\documentclass[12pt,a4paper]{article}

\pdfoutput=1
\usepackage{graphicx}
\usepackage{amssymb}

\newtheorem{defn}{Definition}
\newtheorem{exmp}{Example}
\newtheorem{qry}{Query}

\usepackage{authblk}
\usepackage{amsmath} 
\usepackage{amssymb} 
\usepackage[caption=false]{subfig}
\usepackage{wrapfig}
\usepackage{algorithm}
\usepackage{algcompatible}

\usepackage{algpseudocode}
\algdef{SE}[DOWHILE]{Do}{doWhile}{\algorithmicdo}[1]{\algorithmicwhile\ #1}%

\usepackage{times}
\usepackage{geometry}
\title{Sub-query Fragmentation for Query Analysis  and Data Caching in the Distributed Environment}

\author[1]{Santhilata Kuppili Venkata\thanks{santhilata.venkata@nationalarchives.gov.uk}}
\affil[1]{The National Archives, London, United Kingdom}

\author[2]{Katarzyna Musial\thanks{katarzyna.musial-gabrys@uts.edu.au}}
\affil[2]{Advanced Analytics Institute, School of Computer Science, University of Technology Sydney,
Sydney, Australia}

\date{}

\begin{document}

\maketitle

\begin{abstract}

The world of query-response systems heavily depends on the cloud storage solutions, distributed data transfers and locality of users etc. When data stores and users are distributed geographically, it is essential to organize distributed data cache points at ideal locations to minimize data transfers.   
This leads to the question, \textit{what} data to cache in \textit{which location}.
To answer this, we are developing an adaptive distributed data caching framework that can identify suitable data chunks to cache and move across  a network of community cache locations. This paper details the first step of the process: the sub-query fragmentation technique to fragment data into portable objects.  Evaluation  suggests that sub-query fragments enable distributed learning methods to understand query patterns and association between sub-queries. The sub-query objects can be modelled easily as input dataset to implement machine learning models to assist cache maintenance.

\end{abstract}

\section{Introduction}
\label{S:1}
Nowadays, it is common for organizations to store their data distributed geographically such as cloud storage systems. Organizations provide a query-response layer over the data storage and hide the location details from users. However, when the responses to user queries (requests) are involved with heavy data transfers, the query analysis provide deep insights into the data placement plans. Clouds and hybrid cloud storage systems are in need of newer technologies to tackle the data placement problem.
If the user queries request for a specific piece of data repeatedly, the data item becomes a candidate for caching. Since users are also distributed, a data item can be requested from multiple locations. The data item may be requested repeatedly from various locations. We need to analyse the patterns of these queries to understand the request flow for a cached data item.

In order to answer the above problem, we have developed \textbf{CommCache} - a distributed adaptive community caching framework to aid distributed data storage systems. CommCache accepts consists of a central query analyser to accept user queries and fragments them to model them into patterns. Queries are fragmented using sub-query fragmentation technique. In this paper, the  sub-query fragmentation technique is detailed to identify the most ideal sub-query fragment for data caching. The sub-queries can be modelled as objects to be placed across network of caches. 

Here is a scenario to understand the CommCache environment. It is common for communities of users from different parts of the world to collaborate on projects and access common data stores.  It is observed that  often their requests to retrieve data partially overlap. It means, it is not necessary that the entire data related to a particular query should be cached. Only a part of the query may be is in common with  another query.  It follows that certain data segments are frequently needed across different locations at different times, as specified by the data flow pipeline. We utilise this feature to propose sub-query fragmentation for distributed  data caching.

When a sub-query $\mathnormal{q}_i$ of a query $\mathnormal{Q}$ is a part of several queries, then $\mathnormal{q}_i$ becomes an good candidate for caching.  Ideally each  $\mathnormal{q}_i$ in $\mathnormal{Q}$ is stored on a cache unit at a location near  the data usage. In our research, we aim to achieve the following: (1) model  data component(s) as independent distributed objects,  (2) define query structures to represent the most frequently searched data and location of the data usage and (3) describe operations for  distributed search and retrieval of the cached data. In this paper, we present formal definitions and modeling of data segments and evaluation of the model under various simulated conditions. 

 \textbf{Sub-query fragmentation} (SQF) \cite{O13-3_adassxxv} defines the process of identifying fragments (sub-queries) of a query execution plan. Sub-query fragmentation follows the rules of semantic caching \cite{Dar96, Keller96, Ren03} for initial query fragmentation. Initially, a query plan and result are cached  according to semantic caching rules. But over repeated accesses, query plans are fragmented into sub-queries or aggregated and cached as separate queries. Periodically, the cached queries are examined for their frequency of use. SQF differs from other fragmentation techniques in two ways. One, cache units save  sub-queries that are easily portable across cache units. When multiple users request the same segment from various geographical locations, transferring a smaller data segment consumes fewer network resources. Two, with smaller chunks, the aggregation process of query results do not need to be performed on the data server. Instead, processing can be delegated to the user location, thus  avoiding resource-consuming processing at the data server. It leads to the late binding of partial results to process the query result only when and where it is needed.  SQF facilitates  a quick lookup of frequent sub-queries and  remaining parts of the cached query are evicted when they are not in demand.  Distributing smaller segments to caches near user locations provide support for technologies such as Edge caching and Fog computing \cite{Varghese17, Stojmenovic14, 5GSurvey}. 

\section{Background \& Related Work}
\label{sec:bgrw}
The concept of data caching  to reduce response time and volumes of data transfers is highly researched in the past. Multiple experiments with a  variety of cache grains are available in the literature. Page caching and tuple caching are and the most widely used approaches  to cache results of the query. Page caching  caches a collection of index and data pages \cite{DeWitt90}. Tuple caching caches collections of tuples  \cite{DeWitt90, Ullman}.  Though  of tuple caching  caches accurate data, the maintenance overhead makes tuple caching a non-feasible technique to implement in distributed applications.  Attribute caching is even a fine grained model: it stores data \cite{PapadomanolakisA04} at the attribute level of a table. Unless attribute caches are highly specific to an application, they may not be used for larger tables. Both tuple and attribute caches have very high maintenance for general applications. None of these  caching methods is suitable for distributed caching for the overhead to maintain data from multiple databases. 

Chen et al. \cite{Chungmin94} were the pioneers  of query result caching and query matching.   In their work, authors proposed about intermediate results to cache. They have defined  a directed graph whose nodes reference to base relations and cached temporaries and edges represent derivation paths. Their graph representation helps the direction of query fragments and the path to process the data. 

Unlike pure result caches, semantic caching \cite{Dar96} and predicate-based caching \cite{Keller96} cache queries along with query results. Queries form metadata along with the records of the data.
This approach allows partial matching of a query result even when the cached query results do not satisfy the new query entirely. These methods provide an accurate semantic description of the content of the query. This approach is much better than result caches as the semantic cache allows more queries to reuse the already cached content. It also, allows the formulation of a query to retrieve an exact set of missing result tuples from the server. The semantic caching model introduced concepts called semantic regions: the probe query and the remainder query. Semantic regions within the cache are associated with the collection of tuples. Their replacement policies for maintenance are based on the usage information of semantic regions. A probe query is the part(s) of the query that can be answered by the cache, and the remainder query is the part(s) of the query for which data should be brought from the databases (the cache misses). This initial model of the semantic cache only supports select-project queries but does not support joins in the query. Dar et al. \cite{Dar96} compared their query and result caching with tuple caching, and page caching and show that semantic caching avoids the high overheads of two traditional caching methods. Keller et al. \cite{Keller96} present their model of semantic cache to be able to process select-project-join queries. They allow `WHERE' conditions and range predicates. This caching scheme is for a central server with multiple clients. They claim lower response times and reduced message traffic, higher server throughput and scalability compared to page-caching. However, semantic caching suffers with the addition of attributes or dropping parts from WHERE conditions, which might increase the resultset size. This approach may waste memory resources which are at a premium in caches. Also, it may add to network communication costs. Another issue with semantic caching is, when a cache entry is not enough to answer a new query, the remainder query must fetch the new data and merge it with the existing cache entry. It may include joining the cache results with a new table, extending ranges of \emph{WHERE} clauses and other constraints. The system must take care such that the remainder query and the merging of the remainder query with the cache entry do not cost more than the original new query. However, if the construction of the remainder query is successful, the semantic caching produces little overhead and reduces response time and network overhead \cite{Jonsson06}. Semantic caches have caught the interest of many researchers for their ability to identify semantic meaning. In a further development, Ren et al. \cite{Ren03}  introduced a formal semantic cache model  for select-project queries and single relations. They explained coalescence and decomposition  to avoid redundant data. To use semantic caching in the distributed environment, Ryeng et al. \cite{RyengHN11} proposed a globally distributed cache based on autonomous sites. They have built caches for intermediate results on the sites where the results are produced. In their design, subsequent similar queries can benefit from the stored caches.  Semantic caching has been applied to deductive databases \cite{Chakravarthy90}, federated databases \cite{Goni1997}, web sources \cite{Lee01, Lee99}, and web querying systems \cite{Chidlovskii99}. They are all built for a single entry point to the system. 

Since semantic caching creates a region of interest, it makes sense to cache a join of two tables rather than individual tables themselves. With semantic caching, it is possible to cache query plans and collect statistics for the use of join product of tables rather than tables. In general, semantic caching proves to be an ideal candidate to extend as it keeps the semantic record of a query. However, semantic cache seems to be ill-suited for distributed databases due to the remainder query merging. We need to address this issue and find a suitable way to make it applicable to distributed databases. Though semantic caching can be successfully applied for middleware caching \cite{RyengHN11}, it needs to be extended to match the user query request patterns and cache data placement. A distributed cache model needs to be defined for content caching (query and result caching). 

Another stream of caching is to cache query execution plans. Instead of caching an entire dataset,  smaller join resultsets from execution plans are identified to cache. Since query execution plans depending on the locality of the data, execution plan caching is more suitable for distributed data caching. 
 Rao et al.\cite{Rao:1998}, proposed the invariant technique for correlated queries. In nested queries, sub-queries that are not related to the external references are cached for reuse. We use the idea of independent sub-query or invariants.  Our model uses query execution tree to obtain the sequence of sub-queries to identify the most used part of the query.  Zanfaly et al.  \cite{ZanfalyAE04} performed analysis of multi-level caching of query plans in distributed databases. They have published the distributed model of query caching at pre-defined locations. This model does not include the performance for changing workloads for changing user needs. A detailed study about query language, access methods, cost-based query optimiser are presented for the structured and  semi-structured data \cite{Jason}.

In other types of caching for distributed caching, the object caching method is proposed for data-intensive applications by Haas et al.\cite{Haas99}. Their approach was to load the cache with relevant objects as a by-product of query processing. Due to the applicability with objects,  object caching seems to be an ideal way \cite{Degenaro2000} to implement with middleware systems to access non-relational databases. Object caching gains by having fewer faults by eliminating the need to fragment in objects. The object-based approach seems to be a way to cache contents on the distributed systems.  However, object caching suffers from an increase in the cost of queries incurred by storing extra data. Also, queries may need multiple objects. Hence, the overhead incurred with object-level caching may be more than query caching. Many studies have been carried out on distributed multi-query processing by \cite{Andrade02, Nam10, d'Orazio07} using active semantic caching. Andrade et al. \cite{Andrade02, Nam10} studied the benefit through cached results in the proxy either directly or through transformations  to the results. They exploited processing commonality across a set of concurrently executing queries and reduce execution times by using previously computed results with an objective to produce better query optimisation. 

In an extension, D'Orazio et al. \cite{d'Orazio07} have proposed distributed query scheduling policies in the grid environment. These policies directly consider the available cache contents by employing distributed multidimensional indexing structures and an exponential moving average approach to predict future cache contents. While the aim to re-use partial query results is similar to us, we differ in approach to developing sub-queries. They identify parts of a semantic query and use the coordinates for indexing. Our approach is to divide a user query into sub-queries based on the query plans. Our method allows sub-queries to associate with others when they are repeatedly requested together.  The advantage of sub-query fragmentation approach is, it is possible to transfer the sub-query (as a part of pre-placing) when needed at other cache locations. Our approach makes query caching suitable for distributed computing. 
\section{CommCache - Distributed Cache Environment}
\label{sec:DCE}
\begin{figure}[ht]
\centering
\includegraphics[width=13.7cm, height=6.5cm]{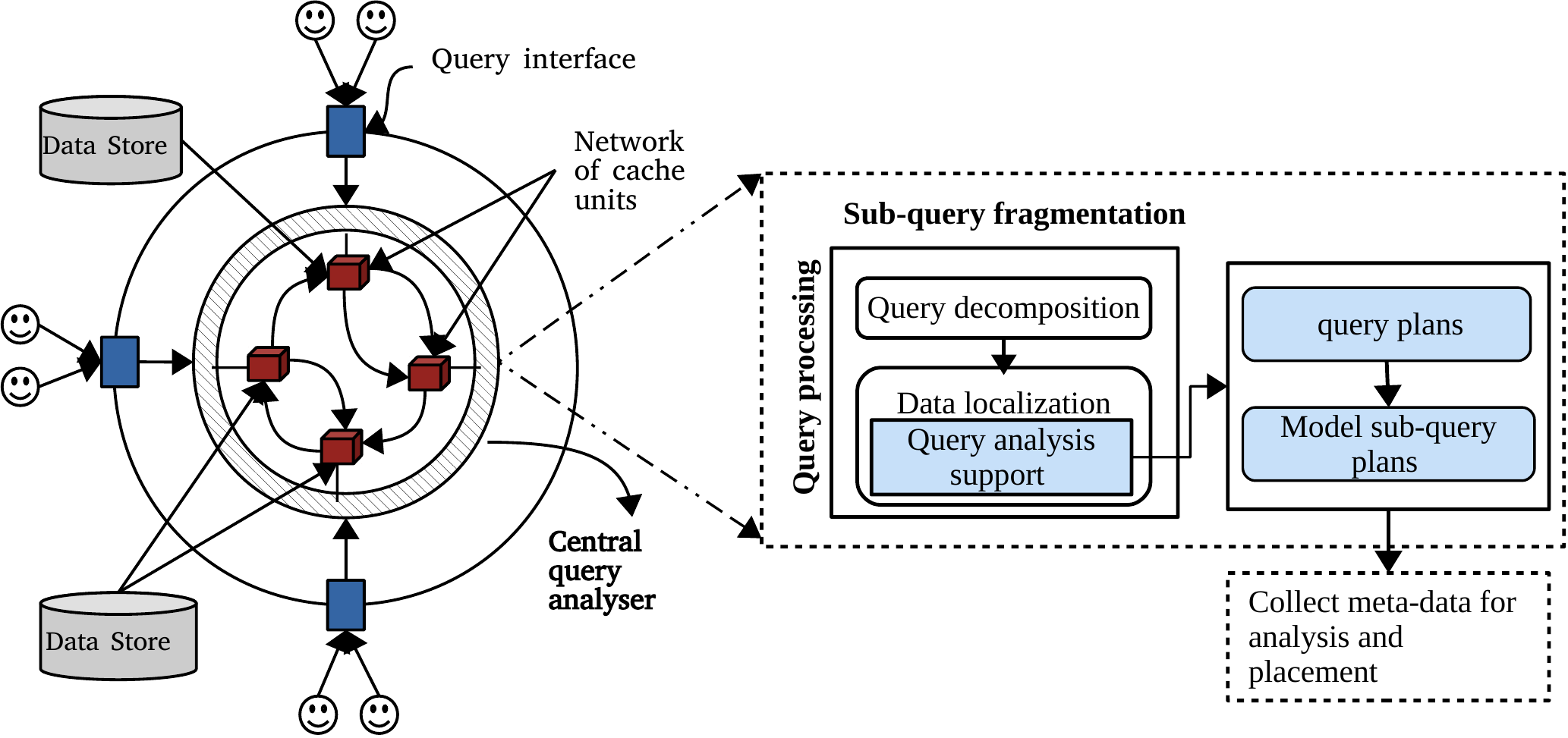}
\caption{ Sub-query fragmentation unit in the CommCache environment}
\label{fig:Distributed_environment}
\end{figure}

This section details the key architectural features for query analysis of CommCache with the help of examples. In its current version, CommCache is implemented with  a global centralised environment for query processing. We consider a network of distributed caches connected to the central query processing unit. As shown in Figure \ref{fig:Distributed_environment}, user queries are submitted to the central query analyser. The query processor receives user queries and makes query execution plans. 

The working of CommCache is divided  into two phases; (i)  the \emph{active cache phase} during which, the cache management collects queries and query patterns are analysed and (ii) the \emph{cache maintenance phase}, the cache replacement policies are chosen to maintain the cache coherence and re-usability of cached content. Both these phases complement each other to decide \textbf{what} to cache, \textbf{how long} to cache and \textbf{where} to cache the data.  The query analyser collates query plans, fragments them into sub-queries and assumes the responsibility for searching and retrieval of all fragments available for a query during the active cache phase. During the maintenance phase, it examines semantic regions of interest and relocates sub-queries according to demographic information,  if necessary. Thus, SQF tracks changing workloads and readjusts the cache contents. 

 In our simulations, the query workloads are assumed to be a continuous stream of queries from users to distributed data stores.
Throughout this paper, the concepts and notions introduced will be illustrated by means of two scenarios  with  sample queries.\\

\begin{figure}[ht!]
 \centering
 \includegraphics[scale=0.5]{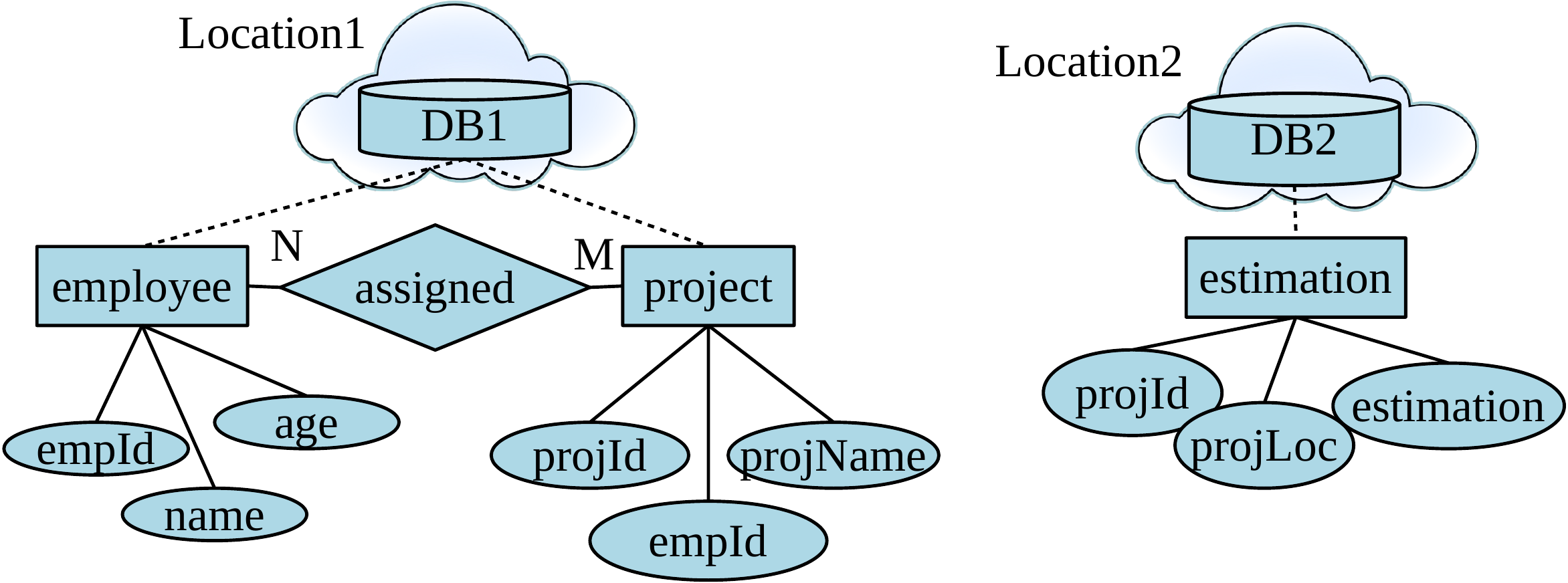}
 \caption{Two databases and their contents used in Scenario 1}
 \label{fig:cloud_ex1}
\end{figure}

\noindent \textbf{Scenario 1.}  There are two  non-replicated  data\-bases (DB1 and DB2) at two locations (separated geographically), \emph{Location1} \& \emph{Location2} respectively. Tables \emph{employee(empId, name, age)}, \emph{project(projId, projName, empId)} are part of \emph{DB1} and table \emph{estimation(projId, projLoc, cost)} is a part of  \emph{DB2}. The database instance is shown in the Figure \ref{fig:cloud_ex1}. These  databases do not need be  homogeneous databases. We assume translation of data across heterogeneous databases to be an implicit step during the query planning. Consider  queries, query\ref{qry:1}, query\ref{qry:2} and query\ref{qry:3} that request data from these databases.

\begin{qry}
This query is to find names of employees, their projects and the project locations, where project cost is under 50,000.
\begin{small}

\noindent{\tt{SELECT empName, projName, projLoc }}\\
{\tt{FROM employee(DB1), project(DB1), estimation(DB2)} }\\
{\tt{WHERE ((employee.empId=project.empId)} AND\\ (project.projId=estimation.projId)}
{\tt{AND \\ (estimation.cost $<$ 50000))}}\\ 
\end{small}

\label{qry:1}
\end{qry}

\begin{qry}
This query obtains the names of projects that cost under 50,000 where employees older than 45 years of age are working.

\begin{small}
\noindent{\tt{SELECT projName}} \\
{\tt{FROM  employee(DB1), project(DB1), estimation(DB2)}} \\
{\tt{WHERE ((employee.age $>$ 45)AND((employee.empId = project.empId)\\AND (estimation.cost $<$ 50000)))}}
\end{small}
\label{qry:2} 
\end{qry}

\begin{qry} This query is to find all employee names who are older than 45 years.

\begin{small}
 \noindent{\tt{SELECT empName}}\\
 {\tt{FROM employee(DB1)}} \\
 {\tt{WHERE (employee.age $>$ 45)}} 
 \end{small}
 \label{qry:3}
\end{qry}

\noindent \textbf{Scenario 2.}
\begin{figure}[ht!]
 \centering
 \includegraphics[scale=0.55]{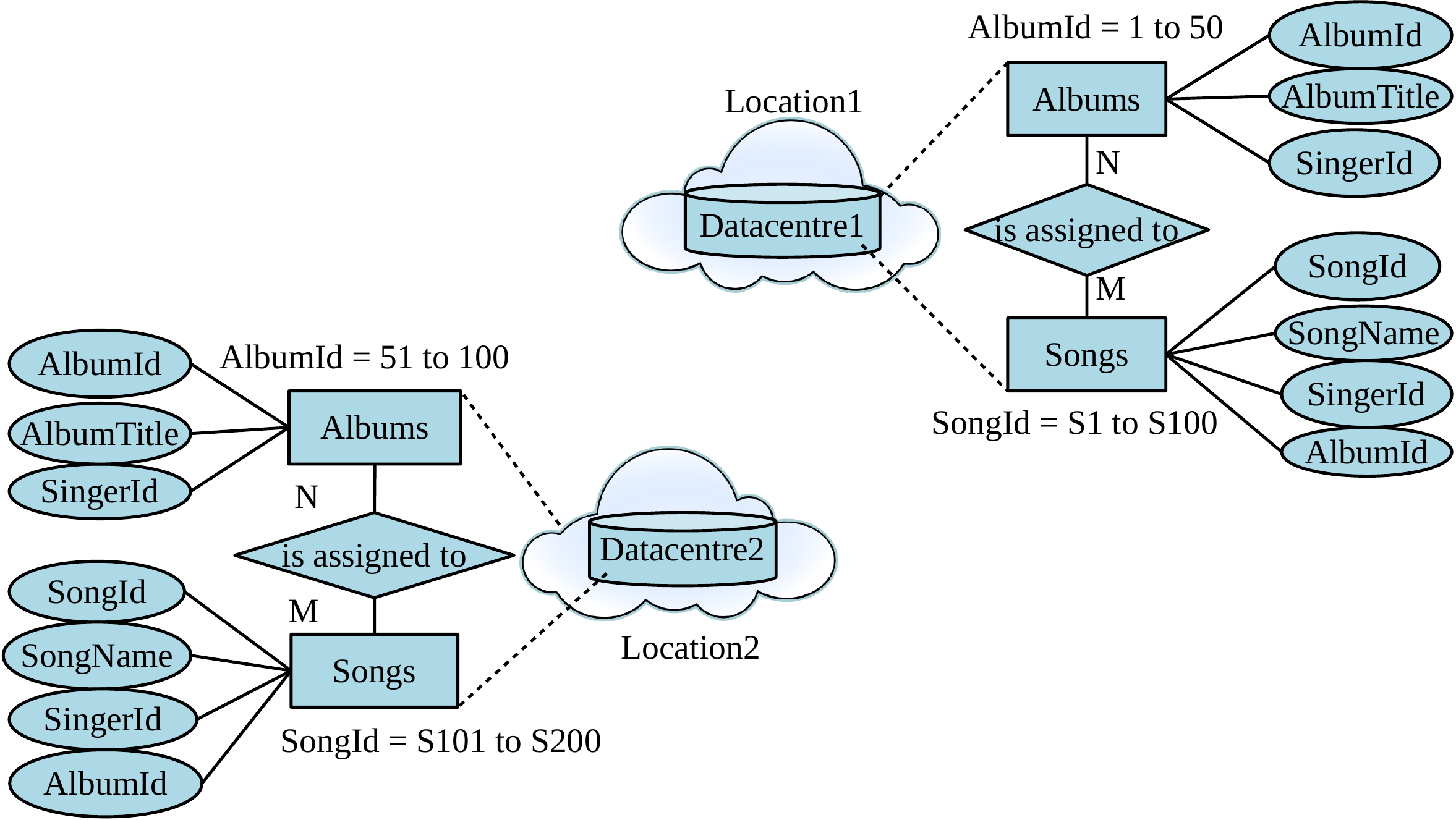}
 \caption{Model of horizontally fragmented distributed data centres in Scenario 2.}
 \label{fig:cloud_ex2}
\end{figure}
 Two horizontally fragmented distributed data centres contain horizontally fragmented data of two tables \emph{Albums(AlbumId, AlbumTitle, SingerId)} and \emph{Songs (SongId, SongName, SingerId)} at two  geographically separated locations. The table 'Albums' is horizontally fragmented on the primary key \emph{'AlbumId'}. Similarly, the table \emph{'Songs'} is horizontally fragmented on the primary key \emph{'SongId'}.

All tuples in \emph{Albums} table ranging from 1 to 50 are stored at  Loc\-ation1 in \emph{Data\-store1} and  tuples from 51 to 100 are stored at \emph{Loca\-tion2} in  \emph{Data\-store2}. 
All tuples in the \emph{Songs} table ranging from S1 to S100  are stored at \emph{Loca\-tion1} in \emph{Data\-store1} and the tuples from S101 to S200 are stored at \emph{Loca\-tion2} in \emph{Data\-store2}. Database instance is as shown in Figure \ref{fig:cloud_ex2}. 
Consider the following queries sent to these data centres\footnote{This scenario is a reproduced  version of the example given at https://cloud.google.com/spanner/docs/query-execution-plans}.

\begin{qry}
This query obtains all song names stored in the table \emph{Songs} from all fragments of the table.

\begin{small}
\noindent{\tt{SELECT SongName FROM Songs;}}
\end{small}
 \label{qry:4}
\end{qry}

\begin{qry}
This query is to get  titles of albums and names of songs sung by singers present in albums.

\begin{small}
\noindent{\tt{SELECT al.AlbumTitle, so.SongName \\
FROM Albums AS al, Songs AS so \\
WHERE al.SingerId=so.SingerId AND al.AlbumId = so.AlbumId}}
\end{small}
\label{qry:5}
\end{qry}

\section{Sub-Query Fragmentation}
\label{sec:SQF}
The previous section introduced the environment within which dispersed groups of users query distributed databases. This section introduces the theoretical model of sub-query fragmentation that enables the distributed caching system that serves those groups of users.  

\subsection{The Sub-query Concept} 
\noindent We can formally define a sub-query as
\begin{defn}
 A query $\mathnormal{{q}_i}$ is a sub-query  of a query $\mathnormal{Q}$ if $\mathnormal{{q}_i}$ is a valid query that can be executed independently and its result can be combined with that of other sub-queries of $\mathnormal{Q}$ to produce the result of $\mathnormal{Q}$.
\end{defn}
 A sub-query can be further fragmented into one or more sub-query plans depending on the need and possibility to fragment further. Since  sub-queries are the parts of query execution plans, an \textbf{atomic sub-query} is defined as an  indivisible sub-query or the aggregation of sub-queries that are often queried together. One of the possible query plans for this query  is given in the Figure \ref{SampleQ1}. In this plan, all the data needed from  DB1 is retrieved through a single access. For the purpose of illustration of distributed processing of the query, this plan ignore any local execution plans within a database to compute local results.

\begin{exmp}
Consider the Query \ref{qry:1} from Scenario 1. 
\begin{figure}[ht]
 \centering 
 \includegraphics[scale=0.7]{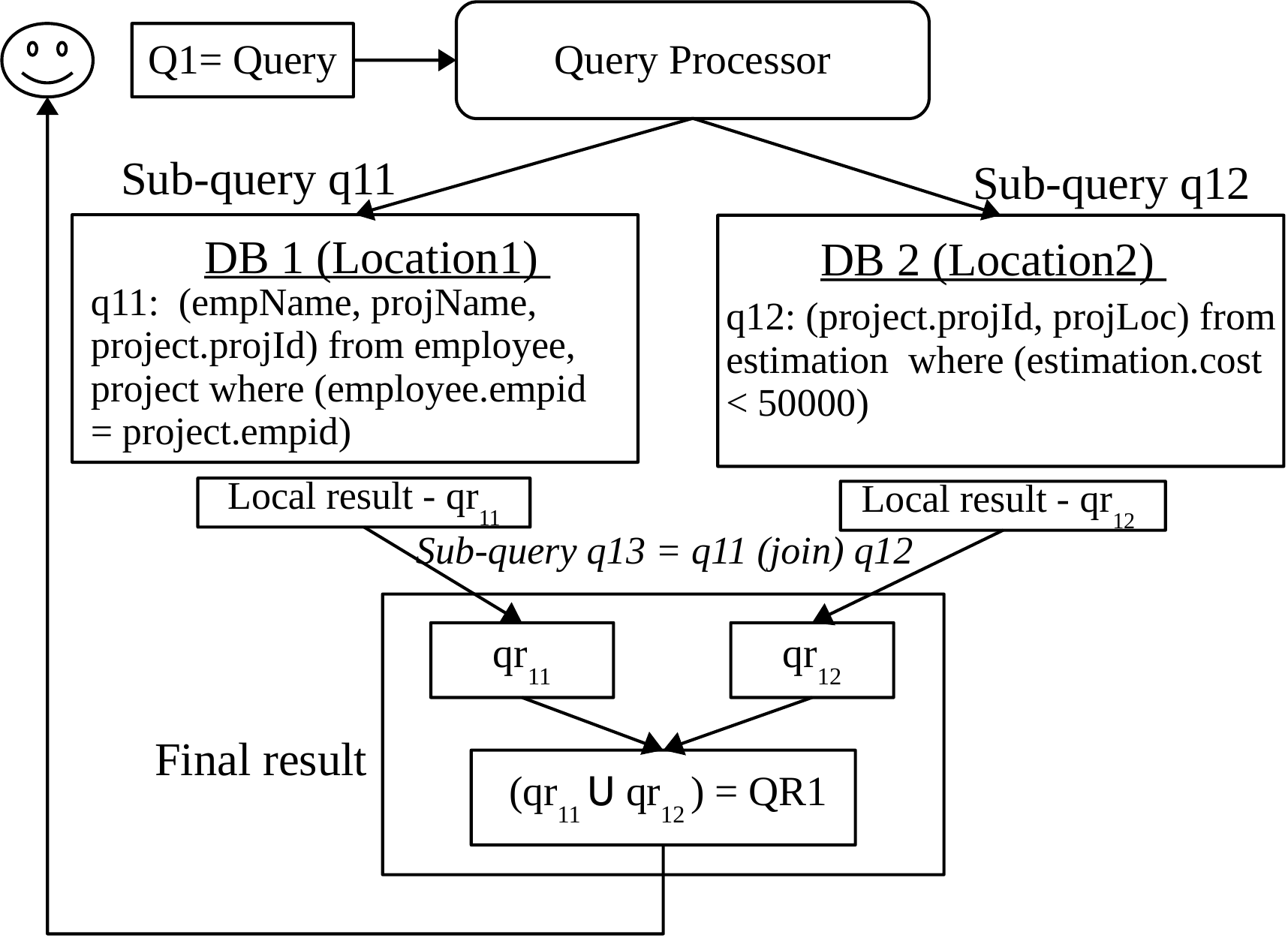}
 \caption{Sample query plan for Query \ref{qry:1}}
 \label{SampleQ1}
\end{figure}

\begin{small}
\begin{itemize} 
\item {\tt{\underline{\emph{sub-query $q_{11}$:}} \\SELECT employee.empId, project.projName, project.projId\\FROM   employee, project\\ WHERE employee.empId = project.empId}} $\Rightarrow$ (partial) local result \emph{($qr_{11}$)}
 
\item {\tt{\underline{\emph{sub-query $q_{12}$:}} \\SELECT (estimation.projId, estimation.projLoc)\\FROM estimation\\WHERE (estimation.cost $<$ 50000)}} $\Rightarrow$ (partial) local result \emph{($qr_{12}$)}

\item {\tt{\underline{\emph{sub-query $q_{13}$:}} \\SELECT (empName,projName, projLoc) \\FROM  $qr_{11}$, $qr_{12}$\\ WHERE ($qr_{11}$.projId = $qr_{12}$.projId)}} $\Rightarrow$ Final result \emph{(QR1)}
\end{itemize} 
\end{small}

\label{ex:sub-query}
\end{exmp}

\noindent The sub-query \emph{$q_{11}$} has a join operation between tables \emph{employee} \& \emph{project}.
Sub-query \emph{$q_{12}$}  is  an \emph{atomic} sub-query, as  the result data fragment  is produced as a single unit. 
Sub-query \emph{$q_{13}$} is a distributed join (or aggregation shown as $\bigcup$ in Figure \ref{SampleQ1}) of the result of two sub-queries \emph{$q_{11}$} and \emph{$q_{12}$}.
Since sub-queries are executed independently, each sub-query can be considered as a query on its own.

In general, a sub-query $\mathnormal{q}_i$  can be described by a tuple $\mathnormal{q}_i$=$\langle \mathnormal{q_i}^R,\mathnormal{q_i}^A,\mathnormal{q_i}^P,\mathnormal{q_i}^C\rangle$ as defined in \cite{Ren03}  and is explained in Table \ref{tab:sub-query}. Since atomic sub-queries are similar to semantic segments, we can derive that a sub-query $\mathnormal{s}$ can be answered by $\mathnormal{t}$ given that ($s^C \wedge t^C) = \mathnormal{s^C}$. In other words,  $s^C\subset t^C$. Similarly, other comparisons such as $s^C$ is \emph{equivalent} to $t^C$, to determine whether $\mathnormal{s}$ can be answered by $\mathnormal{t}$ follow the definitions of \cite{Ren03} while searching the cache for contained query.

\begin{table}[ht]
\centering
\def\baselinestretch{1}\selectfont

\begin{tabular}{p{2cm}p{5 cm}p{5.5cm}}

\hline\hline
\textbf{Attribute} & \textbf{Description} & \textbf{Example value}\\
\hline
$\mathnormal{q_i}^R$ & The set of relations in the query& employee, project \\
\\
$\mathnormal{q_i}^A$ & The set of all attributes to be accessed through select part & employee.empId, project.projName, project.projId, project.empId\\
\\
$\mathnormal{q_i}^P$  & The set of predicates & employee.empId=project.empId\\
\\
$\mathnormal{q_i}^C$ & The resulting tuples of the query & \emph{($qr_{11}$)} \\
\hline
\end{tabular}
\caption{ Tuple attributes of a sub-query for the sub-query \emph{$q_{11}$}.}
\label{tab:sub-query}
\end{table}

\subsection{Notational Representation of a Query}

A  Query  $\mathnormal{Q}$ is an aggregation of its sub-queries executed in a combination of parallel and sequential execution as defined by the query plan. 
We introduce an execution operator `$\parallel$'  for sub-queries that can be executed concurrently. Similarly, `\_' represents a sequential operator for sub-queries that must be executed in a predefined sequence.  Thus, a query plan can be written using the infix notation\footnote{Infix notation is a simple to read notational representation for writing algebraic and logical expressions. Operators are written in-between their operands. Parentheses override operators to resolve precedence and associativity.} of sequential and parallel execution operators written in-between their sub-queries. It follows  that,
\begin{equation}
(\mathnormal{q_1}\cup \mathnormal{q_2}\cup..\cup \mathnormal{q_n}) = \bigcup_{k=1}^\mathnormal{n} \mathnormal{q_k} = \mathnormal{Q}
\end{equation}
where, $\cup$ denotes a parallel or sequential execution operator for aggregation. 
 
 A simplified sequence of operations for sub-query execution  to get the query result is shown in Figure \ref{fig:SubQueryPlan_Q1} for  $\mathnormal{Q1}$. Let the intermediate result generated by the sub-query plan \emph{$q_{11}$} be \emph{$qr_{11}$} and that of sub-query plan \emph{$q_{12}$} be \emph{$qr_{12}$}. The overall result is aggregated as $\mathnormal{QR}1$ on join \emph{($qr_{11}$.projId = $qr_{12}$.projId)}. Sub-query plans \emph{$q_{11}$} and \emph{$q_{12}$} can be executed  as two independent plans in parallel. The aggregation for \emph{$q_{13}$} depends on the results of \emph{$q_{11}$} and \emph{$q_{12}$}. It means, execution of  \emph{$q_{13}$} cannot be performed until the execution of both \emph{$q_{11}$} and \emph{$q_{12}$} are completed. Hence  \emph{$q_{13}$} is executed sequentially after \emph{$q_{11}$} and \emph{$q_{12}$}. Using the notation, $\mathnormal{Q}1$ is:

\begin{equation}
 (((q_{11}) \parallel (q_{12}))\text{ } \_ \text{ }\space(q_{13})) \Rightarrow \mathnormal{Q}1  
\end{equation}

\subsection{The Query Evaluation Tree}

\label{def:QET}

 \begin{figure}[ht]
\centering
\subfloat[Query plan with sub-queries]{\includegraphics[scale=0.8]{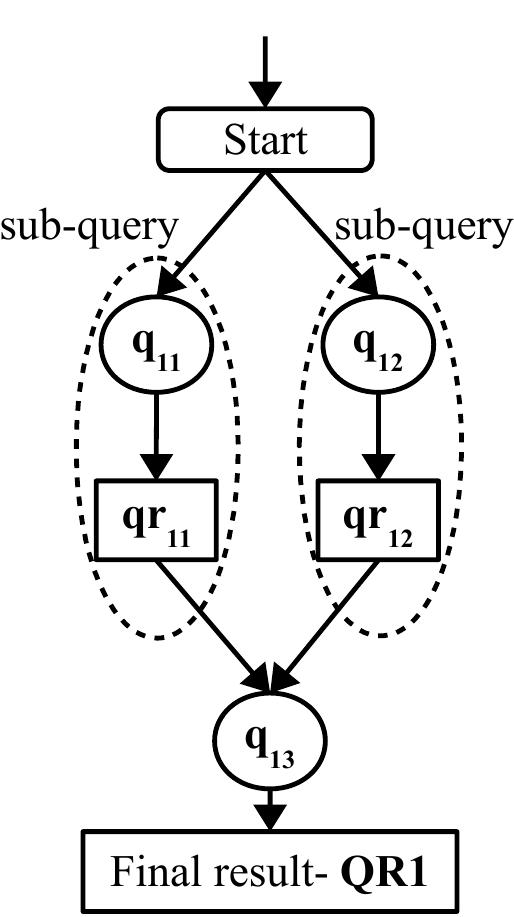}
\label{fig:SubQueryPlan_Q1}}
\hspace{0.1cm}
\subfloat[Tree with execution operators]{\includegraphics[scale=0.65]{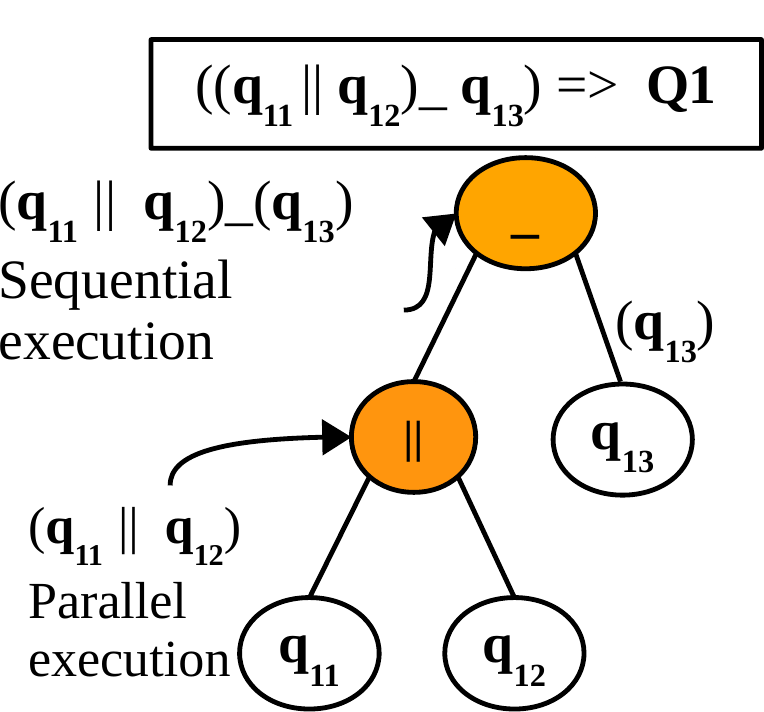}
\label{fig:QET_Q1}}

\caption[]{Query Evaluation Tree (QET) for the query $\mathnormal{Q}1$}
\label{fig:QET}
\end{figure}

A query evaluation tree (QET) for a query $\mathnormal{Q}$  is the graphical representation of a query in the infix order of execution. It is similar to the query execution tree presented in \cite{Rao:1998}. In a QET, sub-queries are found at leaf nodes. Each intermediate node is an execution operation performed on child nodes. When there is more than one child plan to a parent node, it is assumed that child plans are to be executed in the order of left node to right node. When a sub-query needs to be fragmented further, the evaluation tree replaces an execution operator, and further fragmented sub-queries are added as child nodes. 

When a new query appears for the first time, the whole query result is cached as a single node. When parts of the query are requested by other queries, then the original query fragments themselves into sub-queries depending on the demand. Each leaf node of the evaluation tree contains a sub-query plan  and address to the physical location of the sub-query's result. A leaf node consists of metadata of the sub-query result.
The query evaluation tree for the query $\mathnormal{Q}1$ with execution operators is shown in Figure \ref{fig:QET_Q1}.  The structure of the node  is given in Table \ref{tab:node_str}.

\begin{table}[ht]

\centering
\def\baselinestretch{1}\selectfont
\small
\begin{tabular}{p{2.5cm}p{9cm}}
\hline
\textbf{Attribute} & \textbf{Value}\\ \hline
Sub-query ($\mathnormal{q_i}$) &  infix expression (Notational representation of  $\mathnormal{q_i}$)\\  \\

Semantics ($\mathnormal{q_i}$) & $\langle\mathnormal{q_i}^R,\mathnormal{q_i}^A,\mathnormal{q_i}^P,\mathnormal{q_i}^C\rangle$ (Description of semantic segment)\\ \\

Address & a pointer to the physical address of  $\mathnormal{q_i}$ \\ \\
\hline
\end{tabular}
\caption{The  structure of a node in the query evaluation tree}
\label{tab:node_str}
\end{table}

\subsection{Query Complexity}
The Complexity of a query is the number of sub-queries to be executed in a query plan. In other words, complexity is the number of  leaf nodes of the query evaluation tree. 
The complexity of a query is used to understand the processing requirements for a given query. A higher number of sub-queries in a query indicates higher processing overheads. Query complexity influences  outcome of the analysis of query patterns. It is used in the decision making whether it is optimal to fragment a query further during the maintenance phase.

\noindent \textbf{\emph{Example.}} Consider the query \ref{qry:3} from Scenario 1.
This query is to find all employee names who are above 45 years of age. 

\begin{small}
\noindent {\tt{SELECT (empName)}}\\
\noindent {\tt{FROM employee(Database1)}}\\
\noindent {\tt{WHERE (employee.age $>$ 45)}} \\
\end{small}

\begin{figure}[ht]
\centering
\subfloat[Sample query plan for query \ref{qry:3}]{\includegraphics[scale=0.65]{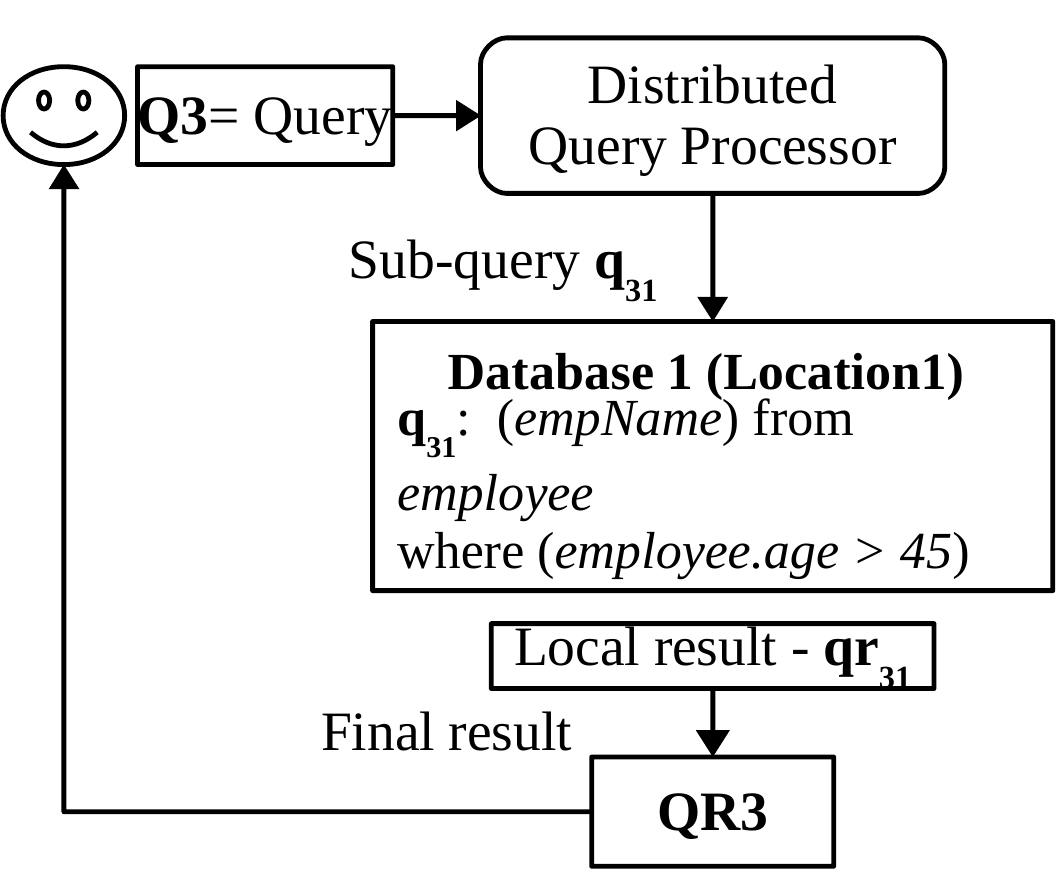}
\label{fig:SampleQ3}}
\hspace{0.10cm}
\subfloat[Tree with execution operators]{\includegraphics[scale=0.7]{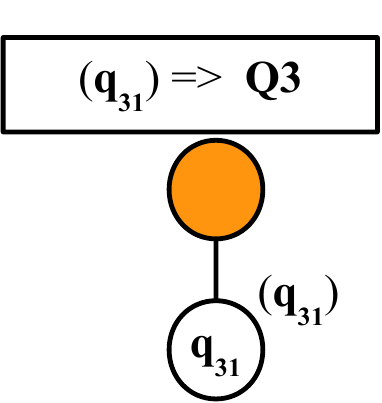}
\label{fig:QET_Q3}}

\caption[]{Query Evaluation Tree (QET) for the query $\mathnormal{Q}3$}
\label{fig:QET3}
\end{figure}
\noindent For this query, a possible query plan is shown in Figure \ref{fig:SampleQ3} with a single sub-query. The equivalent query evaluation tree is shown on Figure \ref{fig:QET_Q3}.

\begin{small}
\begin{itemize}
\item {\tt{\underline{\emph{sub-query $q_{31}$:}} SELECT empName FROM employee(DB1)\\ WHERE (employee.age$>$45)}}
\end{itemize}
\end{small}

\noindent \textbf{\emph{Example.}} Consider  query \ref{qry:4} from Scenario 2.\\
\begin{small}{\tt{SELECT SongName FROM Songs;}} \end{small}
 
\begin{figure}[ht]
\centering
\subfloat[Sample query plan for query \ref{qry:4}]{\includegraphics[scale=0.65]{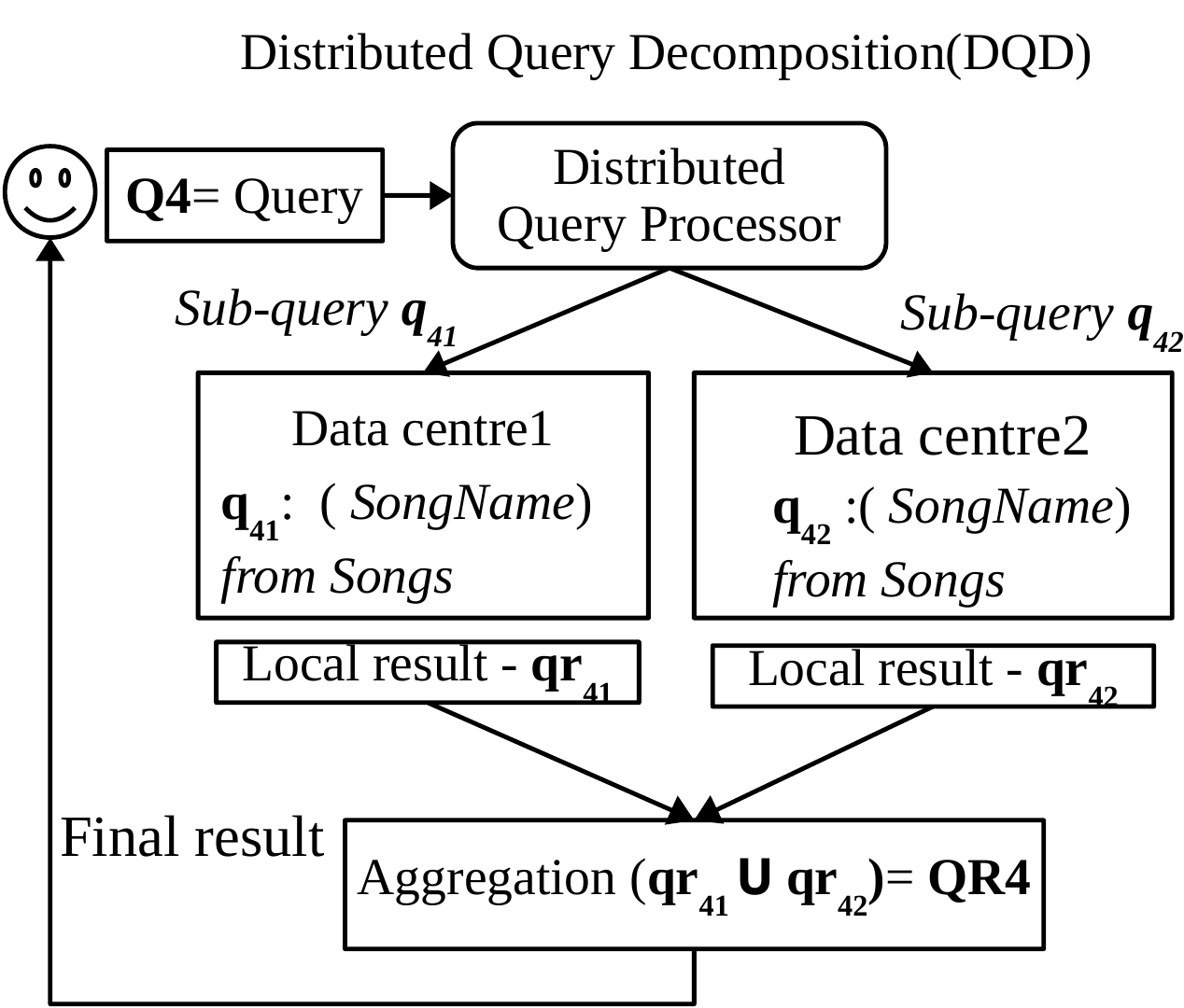}
\label{fig:SampleQ4}}
\hspace{0.01cm}
\subfloat[Tree with execution operators]{\includegraphics[scale=0.65]{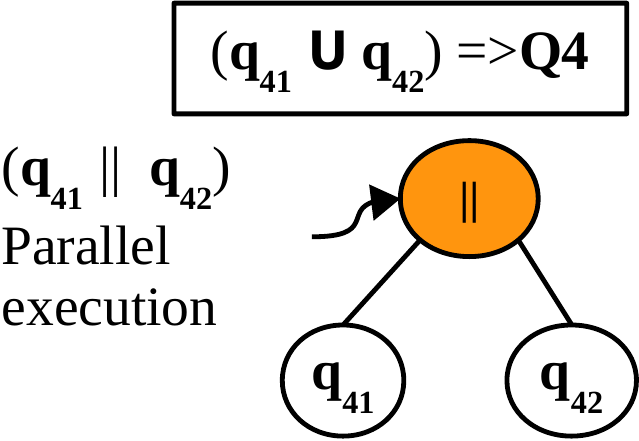}
\label{fig:QET_Q4}}

\caption[]{Query Evaluation Tree (QET) for the query $\mathnormal{Q}4$}
\label{fig:QET4}
\end{figure}

\noindent For this query, a possible query plan is shown in Figure \ref{fig:SampleQ4} with following sub-queries: 
\begin{itemize}
\small
\item {\tt{\underline{\emph{sub-query $q_{41}$:}} SELECT SongName FROM Songs (Location 1)}}\\ 
(This sub-query retrieves all \emph{SongNames} from \emph{SongId=1} to \\ \emph{SongId=100} from the horizontal fragment at Location 1)
\item {\tt{\underline{\emph{sub-query $q_{42}$:}} SELECT SongName FROM Songs (Location 2)}}\\(This sub-query retrieves all \emph{names of songs} from \emph{SongId=101} to \\ \emph{SongId=200} from the Horizontal fragment at Location 2)
\item {\tt{\underline{final result:}  Aggregation of results of sub-query $q_{41}$}\\ and sub-query $q_{42}$}
\end{itemize}
 
Since this query has to access data from two fragments of the table \emph{Songs} from two locations, it is convenient to consider them as two separate sub-queries from the view of network resources. Though the query leads to a semantically single query, it is advantageous to retrieve and cache data from these fragments separately for the sake of the query execution.  Similarly, we consider them as two sub-queries even when the data comes from two replicas of the same table. (As replication is done to do load balancing across servers, it is logical to consider them as two  separate data centres). The third operation is only the aggregation of two query results. Since it is a simple union operation performed on two sub-query resultsets we do not consider as a separate sub-query.
This query has two sub-queries. Hence the complexity of the query \ref{qry:4} is 2. Similarly, the query \ref{qry:1} (in Figure \ref{fig:QET_Q1}) has three leaf nodes. So its complexity is 3. 

\subsection{The Cached Query Model}
\label{subsec:cachedQuery}
When a frequent sub-query is cached, it becomes an independent \emph{cached query object}.
We use a `$\Diamond$' notation to distinguish a cached query  `$\Diamond\mathnormal{S}$' from a user query $\mathnormal{S}$. 

A cached query is made up of two parts. One, that defines the sub-query plan  and second, a pointer to the physical address where the result is cached. The object profile of a cached query consists of  the usage information of the query during  its `cache life time\footnote{The lifetime is the length of time a sub-query is stored in the cache}'. These attributes provide adequate information for the analysis and prediction of future needs of users during cache maintenance.

A \textbf{Cached Query} $\Diamond\mathnormal{S}$  is  a sub-query result and along with its meta-data parameters. A cached query is represented as  a tuple  \\ $\langle\Diamond\mathnormal{S},\mathnormal{CLoc},\mathnormal{V},\mathnormal{C},\{\mathnormal{T}_{uLoc}\},\{\mathnormal{F}_{uLoc}\},\{\mathnormal{D}_i\}\rangle$. \\

\noindent \textbf{\emph{Example.}}  A cached query
  $ \Diamond\mathnormal{S} =\\ \langle((\mathnormal{q}_{a}\parallel\mathnormal{q}_{m})$\_$\mathnormal{q}_{n})$, (cache-2), 10,  3, \{3,4,2\}, \{10,12,10\}, \{$\mathnormal{q}_{a},\mathnormal{q}_{b},\mathnormal{q}_{k}$\}$\rangle$. 
 \\
 The description of these parameters is explained in Table \ref{tab:cachedQuery}.
 
\begin{table}[ht]
\centering
\def\baselinestretch{1}\selectfont
\begin{tabular}{p{1.5cm} p{7cm} p{3cm}}

\hline 
\textbf{Attribute} & \textbf{Description}& \textbf{Value} \\ \hline
$\Diamond\mathnormal{S}$& The query expression of  $\mathnormal{S}$& $((\mathnormal{q}_{a}\parallel\mathnormal{q}_{m})$\_($\mathnormal{q}_{n})$)\\ \\
$\mathnormal{CLoc}$ & Location of the cache server, where $\Diamond\mathnormal{S}$ is currently cached& (\emph{cache-2}) \\ \\
$\mathnormal{V}$ & Volume of the cached data of  $\Diamond\mathnormal{S}$ in GB & 10\\ \\
$\mathnormal{C}$ & Complexity of  $\Diamond\mathnormal{S}$ & 3\\ \\
\{$\mathnormal{T}_{uLoc}$\}& A set of time stamps, $\Diamond\mathnormal{S}$ recently used at each of the user locations 'uLoc'& \{(\emph{uloc-1}),3\}, \{(\emph{uloc-2}),4\}, \{(\emph{uloc-3}),2\} \\ \\
\{$\mathnormal{F}_{uLoc}$\}& A set of frequencies (no.of times query accessed) $\Diamond\mathnormal{S}$ has been requested from each user location 'uLoc' & \{(\emph{uloc-1}),10\}, \{(\emph{uloc-2}),12\}, \{(\emph{uloc-3}),10\} \\ \\
\{$\mathnormal{D}$\} & List of other queries with which $\Diamond\mathnormal{S}$ has been queried together& \{$\mathnormal{q}_{a},\mathnormal{q}_{b},\mathnormal{q}_{k}$\}\\ \\

\hline 

\end{tabular}
\caption{Description of a Cached Query $\Diamond\mathnormal{S}$}
\label{tab:cachedQuery}
\end{table}

\subsection{Cache Granularity}
The \textit{cache granularity}  is the  smallest independent cacheable object. For an attribute or tuple caching policy, the granule is an attribute or a tuple respectively. Commonly, data transfers are measured as the number of memory pages \cite{Silberschatz08}. The physical size of a cache granule is a page.
Though query objects are cached at the page level, from the logical point of view, a sub-query (result) is the granule with sub-query fragmentation.  A SQF cache granule can have a variable size due to the frequency of a sub-query requested along with other sub-queries.

\subsection{Query Equivalence and Overlap}
\label{sec:subsumption}
The query equivalence is essential when checking whether a user query can be answered by the cache. Equivalence checks whether the result needed by a user query is equivalent to,  or part of, any of the cached queries.

\subsubsection{Equivalent Queries}

\noindent \textbf{\emph{Definition.}} Two queries ($\mathnormal{S},\Diamond\mathnormal{T}$) are \textbf{equivalent} $\mathnormal{S}\equiv\Diamond\mathnormal{T}$, when their query evaluation trees have identical internal nodes and the participating sub-queries are  equivalent. It follows that ($\mathnormal{S} \cap \Diamond\mathnormal{T}) = \mathnormal{S}$ or $\Diamond\mathnormal{T}$. \\

\noindent \textbf{\emph{Definition.}} Alternately, two queries ($\mathnormal{S},\Diamond\mathnormal{T}$) are  \textbf{equivalent} $\mathnormal{S}\equiv\Diamond\mathnormal{T}$, when their final results yield the same resultset with different sub-queries.\\

Since the root node of a query evaluation tree is an aggregation of the complete query, equivalence for two queries can be checked by the query  expression  or answerable by the semantic definition \cite{Ren03} (explained in the Table \ref{tab:node_str}) at its root node. \\

\noindent \textbf{\emph{Example.}}
Distributed query plan generation depends  on many factors such as load distribution, process distribution, and location dependence of data availability \cite{Ozsu:2007}. A  query  may have more than one  execution plan  leading to many possible  strategies. For example,  another query execution plan for  Query \ref{qry:1} (from Scenario 1) can be  with following sub-queries:

\begin{small}
\begin{itemize}
\item {\tt{\underline{\emph{sub-query $q_{14}$:}}  SELECT empId,empName \\ FROM employee$\Rightarrow$partial result  ($qr_{14}$).}}
\item {\tt{\underline{\emph{sub-query $q_{15}$:}} \\SELECT project.projName, project.empId, estimation.projLoc\\FROM project, estimation \\ WHERE (project.projId=estimation.projId) AND \\(estimation.cost $<$ 50000)$\Rightarrow$partial result ($qr_{15}$).}}
\item {\tt{\underline{\emph{sub-query $q_{16}$:}} SELECT empName, projName, projLoc\\FROM $qr_{14}$, $qr_{15}$ \\ WHERE ( $qr_{14}$.empId = $qr_{15}$.empId )$\Rightarrow$complete result (QR1).}}
\end{itemize}
\end{small}

\noindent The complete query plan with the above sub-queries is
\begin{equation}
(((q_{14}) \parallel (q_{15}))\text{ } \_ \text{ }\space(q_{16})) \Rightarrow \mathnormal{Q}1 
\end{equation}

\begin{figure}[ht]   
\centering
\subfloat[Execution plan 1]{\includegraphics[scale=0.85]{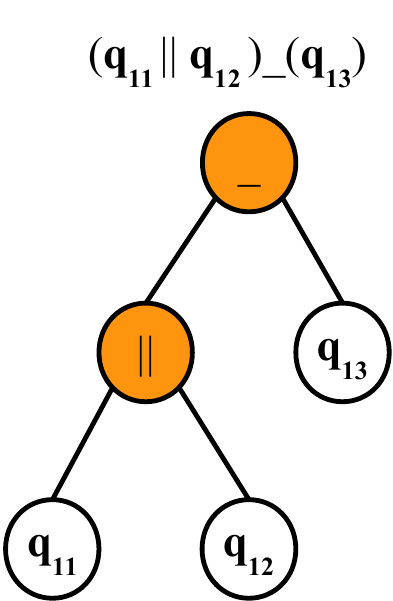}
\label{fig:QET_Q11}}
\hspace{1cm}
\subfloat[Execution plan 2]{\includegraphics[scale=0.85]{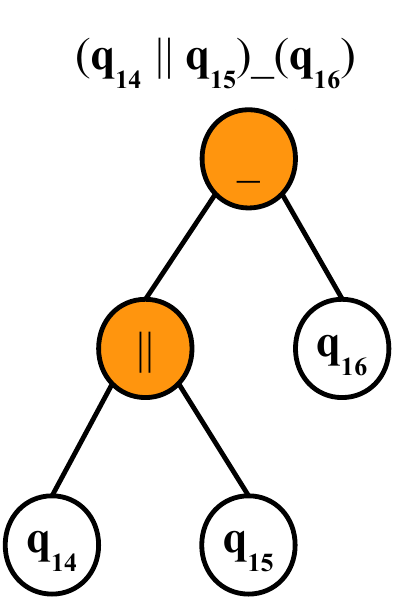}
\label{fig:QET_Q12}}

\caption[]{Equivalent query evaluation trees for $\mathnormal{Q1}$ with different sub-queries}
\label{fig:QET2}
\end{figure}
Two query evaluation trees for $\mathnormal{Q1}$ with sub-queries \emph{$q_{11}$, $q_{12}$, $q_{13}$} and \emph{$q_{14}$, $q_{15}$, $q_{16}$} are shown in \ref{fig:QET_Q11} and \ref{fig:QET_Q12} respectively in the Figure \ref{fig:QET2}. Though,  the leaf nodes of the above evaluation trees are different, the semantic information at the root nodes yields the same resultset, both trees are considered to be \emph{equivalent}. Algorithm to check the equivalence of two queries is given in Algorithm \ref{Alg:QET_Equal}.

\begin{algorithm}[ht]
\small
 \caption{EquivalentQuery ($\mathnormal{S}$,$\mathnormal{T}$)}
  \begin{algorithmic}[1]
  \State \textbf{Input:} Query $\mathnormal{S}$, Query $\mathnormal{T}$
   \State \textbf{Output:} boolean (true or false)   
	\State root-S $\Leftarrow$ root of QET for $\mathnormal{S}$\Comment root node of the cached query
	\State root-T $\Leftarrow$ root of QET for $\mathnormal{T}$\Comment root node of the query to search
	\If {(root-S.sub-query \emph{equals} root-T.sub-query) }
	  \State return \emph{true}
	  \Else { return \emph{false}}
	  \EndIf 
	  
  \end{algorithmic}  
  \label{Alg:QET_Equal}
 \end{algorithm}

\subsubsection{Query Overlap}

\noindent \textbf{\emph{Definition.}} Query $\mathnormal{S}$ is \textbf{completely overlapped} by a query  $\Diamond\mathnormal{T}$, i.e.,  ($\mathnormal{S}\subseteq\Diamond\mathnormal{T}$), when the \emph{root node} of query $\mathnormal{S}$ is \textbf{equivalent} to \emph{one of the nodes}  of the (query  $\Diamond\mathnormal{T}$).  It follows that, ($\mathnormal{S} \cap \Diamond\mathnormal{T}) = \mathnormal{S}$. \\

\noindent \textbf{\emph{Example.}}
Consider  Query \ref{qry:2} and Query \ref{qry:3} from Scenario 1: \\ 
Query \ref{qry:2} obtains the names of projects (with cost $<$ 50000) where employees above the age of 45 years are working.

\begin{small}
\noindent{\tt{SELECT (projName)}}\\ 
 {\tt{FROM  employee(DB1), project(DB1), estimation(DB2)}} \\
\noindent {\tt{WHERE ((employee.age$>$45) AND ((employee.empId=project.empId) \\AND (estimation.cost $<$ 50000)));}}\end{small}\\

\noindent Query \ref{qry:3} is to find all employee names whose are above 45 years of age.

\begin{small}
\noindent {\tt{SELECT (empName)}}\\ 
{\tt{FROM employee(DB1)}}\\  
\noindent {\tt{WHERE (employee.age $>$ 45)}} \end{small} \\

\noindent In the above queries, Query \ref{qry:3} is completely overlapped by Query \ref{qry:2}.\\

\noindent \textbf{\emph{Definition.}}
Query $\mathnormal{S}$ and query  $\Diamond\mathnormal{T}$ are \textbf{partially overlapped},  when \emph{one or more nodes} of the evaluation tree of $\mathnormal{S}$ are \textbf{equivalent} to \emph{one or more nodes } of the evaluation tree of query  $\Diamond\mathnormal{T}$. It follows that, $\mathnormal{S} \cap \Diamond\mathnormal{T} \neq $ \emph{null}. \\

\noindent \textbf{\emph{Example.}}
Consider  sub-queries of $\mathnormal{Q1}$ and $\mathnormal{Q2}$ of Scenario 1.

\begin{itemize} 
\small
\item {\tt{\underline{\emph{sub-query $q_{11}$:}} \\SELECT employee.empId, project.projName, project.projId\\FROM   employee, project\\WHERE employee.empId=project.empId}}
\item {\tt{\underline{\emph{sub-query $q_{12}$:}} SELECT estimation.projId, estimation.projLoc\\ FROM estimation \\ WHERE estimation.cost $<$ 50000}}
\item {\tt{\underline{\emph{sub-query $q_{13}$:}} SELECT empName, projName, projLoc \\FROM  $qr_{11}$, $qr_{12}$\\ WHERE ($qr_{11}$.projId=$qr_{12}$.projId )$\Rightarrow$complete result\emph{(QR1)}}}
\end{itemize} 

\noindent Following are one of the possible query execution plans for Query \ref{qry:2}:

\begin{itemize}
\small
\item {\tt{\underline{\emph{sub-query $q_{21}$:}}  SELECT project.projName, project.projId \\FROM employee, project \\WHERE (employee.empId=project.empId) AND \emph(employee.age$>$45)}}
\item {\tt{\underline{\emph{sub-query $q_{22}$:}} SELECT estimation.projId\\ FROM estimation \\ WHERE (estimation.cost $<$ 50000) }}
\item {\tt{\underline{\emph{sub-query $q_{23}$:}} SELECT projName \\FROM  $qr_{21}$, $qr_{22}$\\ WHERE ($qr_{21}$.projId = $qr_{12}$.projId) $\Rightarrow$  complete result\emph{(QR2)} }}
\end{itemize}

From the above  queries, the \emph{sub-query $q_{22}$} can be answered by   \emph{sub-query $q_{12}$}. If the results of query $\mathnormal{Q}$\ref{qry:1} were cached, then the query $\mathnormal{Q}$\ref{qry:2} can be partially answered by the query $\mathnormal{Q}$\ref{qry:1}. The remaining part of the query $\mathnormal{Q}$\ref{qry:2}, \emph{sub-query $q_{21}$} cannot be answered by the query $\mathnormal{Q}$\ref{qry:1}.

\subsection{Searching for Contained Query}

The process of searching whether a query contains the result is done by checking the equivalence. Since we consider only sub-queries, we introduce a logical layer (such as an index) between the query interface and the physical storage location. The index consists of evaluation trees ordered by the demand for a sub-query. A user query should be searched against each of the root nodes of the cached queries to find what part of the user query can be answered by each of the cached queries. 

When a cached query ($\Diamond\mathnormal{T}$) contains the user query ($\mathnormal{S}$), \emph{i.e.},  $\mathnormal{S}$ is partially or fully overlapped by $\Diamond\mathnormal{T}$, it is enough to search the query evaluation tree of  $\Diamond\mathnormal{T}$ according to breadth-first order since sub-queries are stored in a top-down manner to get the biggest part that can be answered by $\Diamond\mathnormal{T}$. 
We say, set of all sub-queries of  $\mathnormal{S}$ that can be answered by  $\Diamond\mathnormal{T}$ is the \textbf{ContainedQuery  ($\blacktriangledown\mathnormal{S}$)},  
and set of all sub-queries that cannot be answered is the  \textbf{RemainderQuery ($\triangledown\mathnormal{S}$)}.
 Algorithm to search for a completely overlapped is present in Algorithm \ref{Alg:subset}. Algorithm \ref{Alg:contain1} explains to search cache for a user query $\mathnormal{S}$ in general.
 
\begin{equation}
\text{ContainedQuery }(\blacktriangledown\mathnormal{S}) = \{\mathnormal{s}_i \mid \mathnormal{s}_i \subseteq \Diamond\mathnormal{T}\}
\end{equation}
\begin{equation}
\text{RemainderQuery }  (\triangledown\mathnormal{S}) = \{ \mathnormal{s}_i \mid \mathnormal{s}_i \in (\mathnormal{S} - \blacktriangledown\mathnormal{S}) \}
\end{equation}

\noindent From these definitions, we can derive how a  query can be answered by the cache: 

\begin{itemize}
 \item A  query $\mathnormal{S}$ can be \textbf{completely answered} by the  $\Diamond\mathnormal{T}$ in cache if,  $\mathnormal{S}\equiv\Diamond\mathnormal{T}$ or $\mathnormal{S}\subset\Diamond\mathnormal{T}$. 
 \item A  query $\mathnormal{S}$ can also be \textbf{completely answered} by the cache, if all nodes in  $\mathnormal{S}$  can be completely answered by one or more queries in the cache ($\mathnormal{S}=\blacktriangledown\mathnormal{S}$) and ($\triangledown\mathnormal{S} = \varnothing$ (null)).  
 \item A  query $\mathnormal{S}$ can be \textbf{partially answered} by the cache, if  the RemainderQuery  ($\triangledown\mathnormal{S} \neq \varnothing$ (null)). 
 \item A query $\mathnormal{S}$ is \textbf{not found} in the cache if, ($\mathnormal{S} = \triangledown\mathnormal{S}$).
\end{itemize} 

\begin{algorithm}[ht]

  \caption{isContained ($\mathnormal{S} \subseteq \Diamond\mathnormal{T}$)}
  \begin{algorithmic}[1]
      \State \textbf{Input:} $\mathnormal{S}$ 
      \State \textbf{Output:} Internal node ($\mathnormal{t}_i$ )
      \If {EquivalentQuery($\mathnormal{S}$,$\mathnormal{T}$)} 
	  \State return \emph{root-T} \Comment equivalent trees
      \Else {  do Breadth-First-Traversal (root-T)}
	    \If {root-S \emph{equals} internal-node ($\mathnormal{t}_i$)}
	    \State return $\mathnormal{t}_i$
	    \Else { return \emph{null}}
	    \EndIf  
     \EndIf 
  \end{algorithmic}
  \label{Alg:subset}
\end{algorithm} 
 
\noindent \textbf{\emph{Example.}} Suppose we have a cache with 3 queries stored.  A logical view of the cache is shown in Figure \ref{fig:QueryContainment}. Consider a set of queries cached and set of user queries given below. 

\begin{table}[ht]
    \centering
    \begin{tabular}{|c|c|}
    \hline
       \textbf{Cached queries}  & \textbf{User queries} \\ \hline
        $\Diamond\mathnormal{T}_1 \Leftarrow $ ((q1 $\parallel$ q2) \_ (q3)) & $\mathnormal{S}_1 \Leftarrow $ (q1  $||$ q2 )\\
        $\Diamond\mathnormal{T}_2  \Leftarrow $  ((q4) $\parallel$ (q5)  $\parallel$  (q6)) & $\mathnormal{S}_2 \Leftarrow $ ((q3) \_ (q4))\\
        $\Diamond\mathnormal{T}_3 \Leftarrow$ ((q6) $\parallel$ (q9)) & $\mathnormal{S}_3 \Leftarrow $ ((q9) $||$ (q8))\\
        &$\mathnormal{S}_4 \Leftarrow $ ((q7) \_ (q8))\\
        \hline
    \end{tabular}
   
\end{table}

\begin{figure}[]
 \centering
 \includegraphics[scale=0.7]{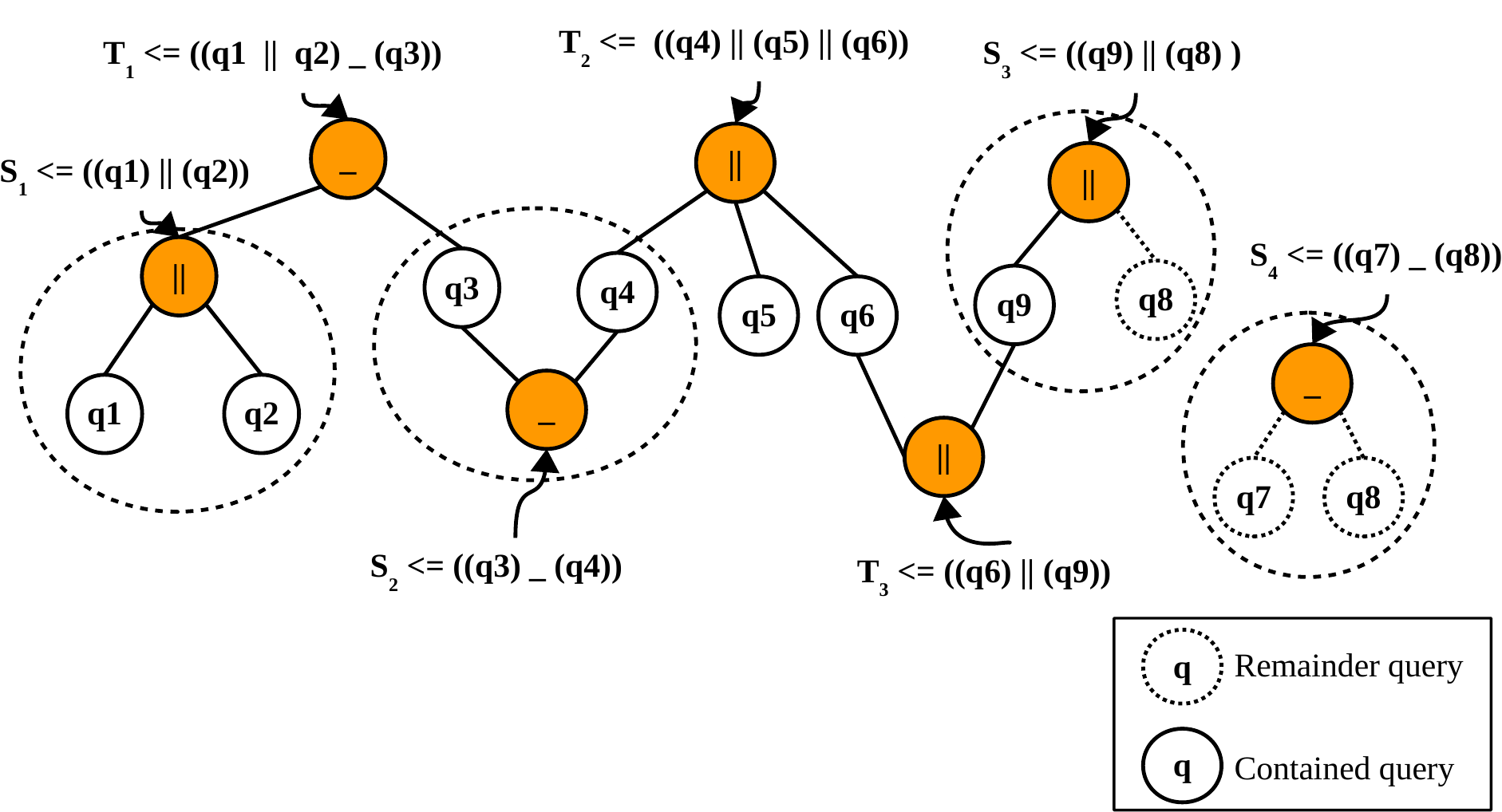}
 \caption{Contained queries and remainder queries}
 \label{fig:FullQueryContainment}
\end{figure}

The logical view of contained and remainder queries for these user queries are shown in Figure \ref{fig:FullQueryContainment}. In this example, $\mathnormal{S}_1$ can be fully answered by the cache as  $\mathnormal{S}_1 \equiv \Diamond\mathnormal{T}_1$. Similarly, $\mathnormal{S}_2$ can also be fully answered by the cache as the leaf nodes (q3) and (q4) of  $\mathnormal{S}2 $ are partially contained in $\Diamond\mathnormal{T}_1$ and $\Diamond\mathnormal{T}_2$. However, $\mathnormal{S}_3$ can be only partially answered by the cache. It has a remainder query $\triangledown\mathnormal{S}_3$ = (q8). The contained query  $\blacktriangledown\mathnormal{S}_3$= q9. $\mathnormal{S}_4 $ cannot be answered at all by the cache as (q7) and (q8) are not stored in cache. Hence, $\blacktriangledown\mathnormal{S}_4 = \varnothing$ and $\triangledown\mathnormal{S}_4 = \mathnormal{S}_4$.

\begin{figure}[]
 \centering
 \includegraphics[scale=0.8]{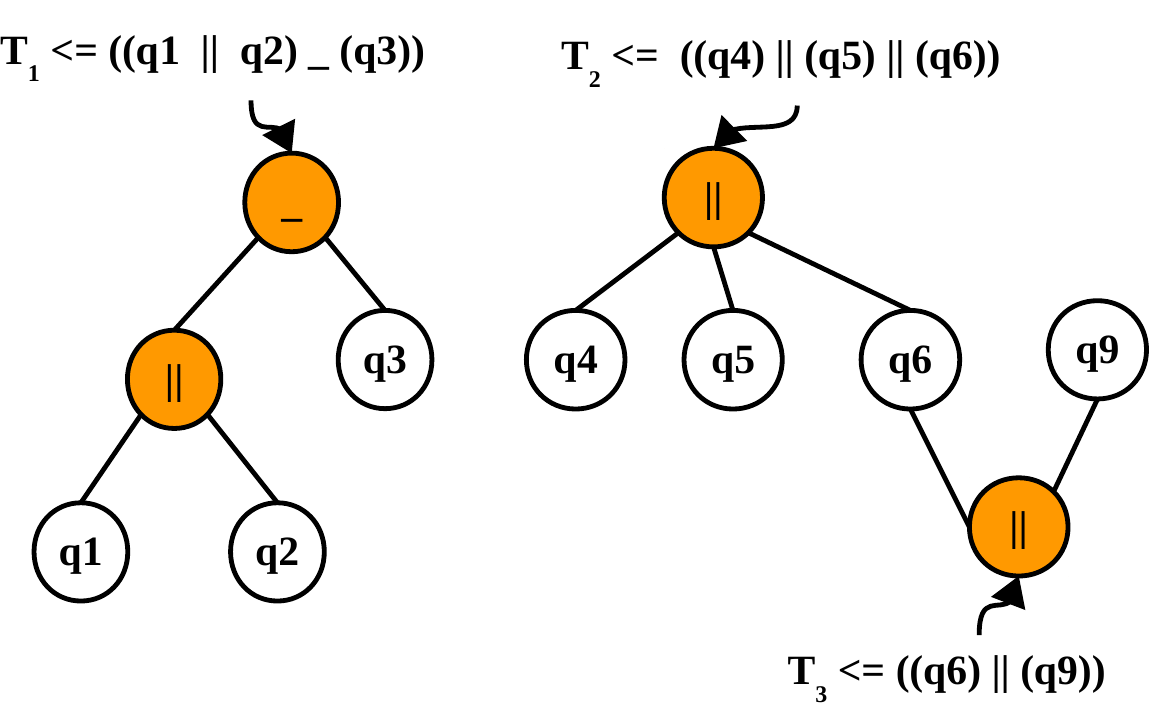}
 \caption{Sample queries present in  cache}
 \label{fig:QueryContainment}
\end{figure}

 \begin{algorithm}[]
 \caption{SearchCache($\mathnormal{S}$) }
  \begin{algorithmic}[1]
  \State \textbf{Input:} Query $\mathnormal{S}$
   \State \textbf{Output:} ContainedQuery  ($\blacktriangledown\mathnormal{S}$), RemainderQuery ($\triangledown\mathnormal{S}$) , status 
    \State $\blacktriangledown\mathnormal{S} = \varnothing$ 
     \State $\triangledown\mathnormal{S} = \{\mathnormal{s}_i \mid \mathnormal{s}_i\in \mathnormal{S} \}$
     \For { each cache unit}
     \For { each cached query $\Diamond\mathnormal{T}$}
     
     ($\blacktriangledown\mathnormal{S},\triangledown\mathnormal{S} $) = \Call{QuerySearch}{$\blacktriangledown\mathnormal{S}, \triangledown\mathnormal{S}, \Diamond\mathnormal{T}$ } 
          \If {($\triangledown\mathnormal{S} =\varnothing$)}
     \State return ($\blacktriangledown\mathnormal{S},\varnothing$, \emph{fully found})     
     \EndIf
   \EndFor  
   \EndFor
   
   \If {($\triangledown\mathnormal{S}$ = $\mathnormal{S}$ )}
   \State return ($\varnothing, \triangledown\mathnormal{S}$, \emph{Not found})
   \Else { return ($\blacktriangledown\mathnormal{S}$, $\triangledown\mathnormal{S}$ , \emph{partially found} )}
   \EndIf
   \State
   \State \%Function to search for a query $\mathnormal{S}$ within a cached query $\Diamond\mathnormal{T}$
   \Function {QuerySearch}{$\blacktriangledown\mathnormal{S}, \triangledown\mathnormal{S}, \Diamond\mathnormal{T}$ }   
  \State \textbf{Output:} ContainedQuery  ($\blacktriangledown\mathnormal{S}$), RemainderQuery $(\triangledown\mathnormal{S})$ 
      \For {each $\mathnormal{s}_i \in \triangledown\mathnormal{S}$}
      \If  {($\mathnormal{s}_i$ $ \equiv \Diamond\mathnormal{T}$) OR ($\mathnormal{s}_i$ $ \subseteq \Diamond\mathnormal{T}$)}
      
      \State $\blacktriangledown\mathnormal{S} \rightarrow \blacktriangledown\mathnormal{S} \cup \mathnormal{s}_i$
      \Else { $\triangledown\mathnormal{S} \rightarrow \triangledown\mathnormal{S} \cup \mathnormal{s}_i$}
      \EndIf
     \EndFor
    \State return ($\blacktriangledown\mathnormal{S}$,$\triangledown\mathnormal{S}$)
 \EndFunction

  \end{algorithmic}
  \label{Alg:contain1}
 \end{algorithm}

\section{Evaluation}
\label{sec:eval}
The overall performance of a cache system depends on active cache phase and maintenance phase together.  
Since the sub-query fragmentation is developed for the distributed cache scenario, we compare and observe the significant improvements using this technique with the baseline full query caching and distributed semantic caching in  distributed environments. 

\subsubsection*{Query Workloads} 
The query workloads are a continuous stream of queries submitted. Queries access data from multiple databases.

sub-query fragmentation needs constantly changing workloads with partial overlaps. We have developed a synthetic query generator  \emph{Qgene} - to generate workloads as query plans with the details of user location(s), cache location and the timestamp.\footnote{ A detailed description of design and query modeling parameters  will be provided for evaluation and use  on request.}. Qgene allows users to set configuration settings for (i) the duration of the observation time window, (ii) number of queries in the workload per window, (iii) varous statistical distributions  for sub-query overlap and inter-query arrival time. Qgene also allows vary query complexity (number of sub-queries). A random factor is introduced to the workloads to eliminate any bias.  Each of the following experiments was executed multiple times and obtained an average value. 

\subsubsection*{Experimental Settings}
All experiments for the evaluation are conducted on  Java based simulator (JDK 1.8). All experiments were conducted for Poisson, Uniform and Exponential distributions of workloads for query overlap and sub-query repetition. 

Other variable settings relate to the cache environment. The maximum number of cache agents is decided based on the inter-cache communications that can be handled by the hardware configuration in the laboratory.  We set the number of cache units in the network to be 20. In the distributed environment, the ideal data placement algorithm to place data near users influence the response time. We set the data placement to follow  Greedy placement policy  for  a uniform evaluation of the caching techniques.  

To understand the impact of cached segments as distributed independent objects and the need for SQF,  the caching techniques  are evaluated and compared for three metrics; (i) average response time, (ii) cache utilisation  and (iii) cache stability. The response time depends on several factors such as (a) caching policies used, (b) cache replacement heuristics, and the (c) distributed data placement methods etc..  Hence, in a fundamental scenario without data placement, SQF and Semantic caching are expected to perform almost the same.  Cache utilisation and cache stability metrics actually highlight the advantage of SQF. Each of these metrics are observed for 14 continuous time epochs. Observations for every epoch were repeated for 8 to 12 times. The average is plotted with standard errors. 

\subsection{Average Response Time}
\begin{table}[ht]
\centering
\begin{tabular}{p{2cm}p{8.5cm}}
\hline
\textbf{Notation}& \textbf{Description}\\ \hline
N & total number of queries \\ 
$\mathnormal{l}_t$  &  average cache latency \\
$\mathnormal{D}_Q $& processing time for a query $\mathnormal{Q}$ at data servers \\
$\mathnormal{Q}_{proc} $& average processing time for a query $\mathnormal{Q}$ \\
$\mathnormal{t}_{qi}$  & data transfer time on network for $\mathnormal{q}_i$ from cache \\
$\mathnormal{V}^{not found}_{Q}$ & volume of data not found in cache \\

\hline
\end{tabular}
\caption{Notation Table}
\label{tab:notation}
\end{table}
The response time is measured as the time elapsed from a query sent from the user to the time the reply is received. It is a relevant metric  as the objective for any caching technique is to minimise the response time. For continuous query workloads, we consider the  average response time per workload.   In the simulation environment, the response time is measured as logical clock ticks.  The symbol notation to calculate average response time lapsed is presented in the Table \ref{tab:notation}. 

 \begin{equation}
\text{Average response time}  =  \frac{1}{N}  (\mathnormal{l}_t +  \mathnormal{D}_Q + \mathnormal{t}_{Q} + \mathnormal{Q}_{proc} + \mathnormal{t}_{qi} )
 \end{equation}
 
 \noindent Where, \\
$\mathnormal{l}_t$ =  query lookup time + actual data retrieval time \\  
$\mathnormal{D}_Q$ = $\mathnormal{V}^{not found}_{Q}$ * $\mathnormal{D}_t$\\ 
$\mathnormal{t}_{Q}$  = time to transfer $\mathnormal{V}^{not found}_{Q}$ *  $\mathnormal{d}_{net}$ \\      
$\mathnormal{t}_{qi}$  = Inter cache transfer time * number of hops

\begin{figure}[ht]

\begin{minipage}{.5\linewidth}
\centering
\subfloat[Poisson query input]{\label{fig:ResponseTime_Granularity_Poisson}\includegraphics[scale=.6]{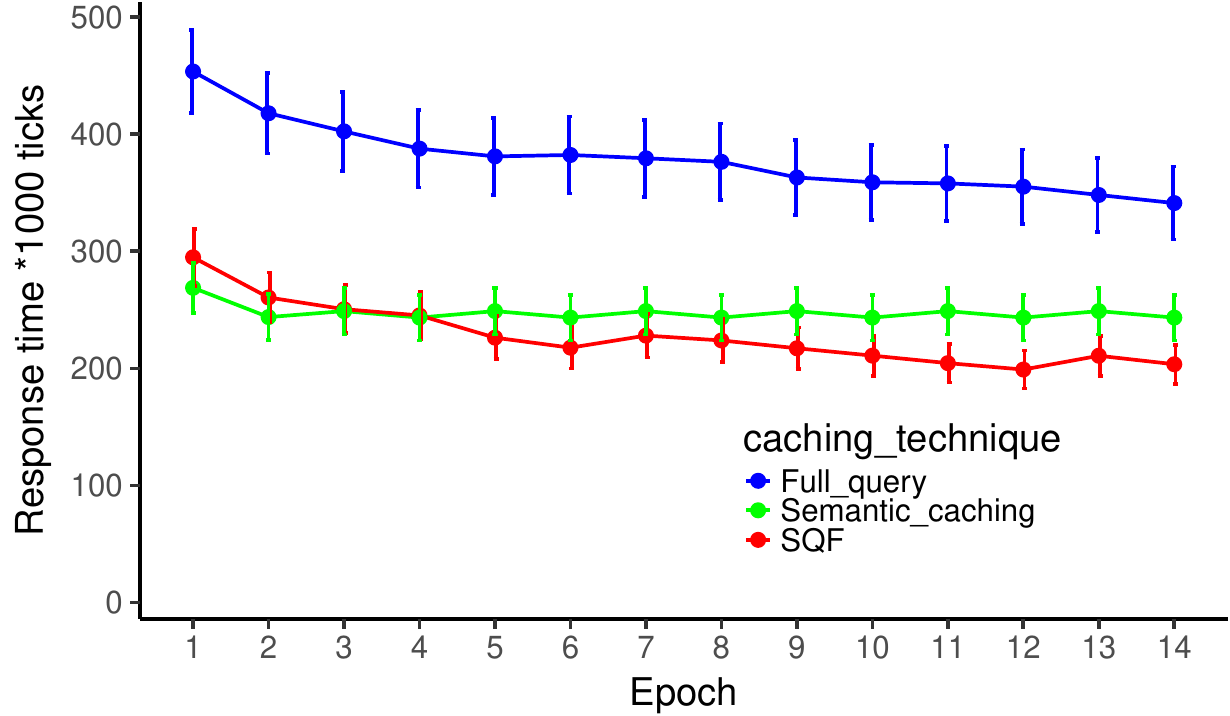}}
\end{minipage}%
\begin{minipage}{.5\linewidth}
\centering
\subfloat[Exponential query input]{\label{fig:ResponseTime_Granularity_Exponential}\includegraphics[scale=.6]{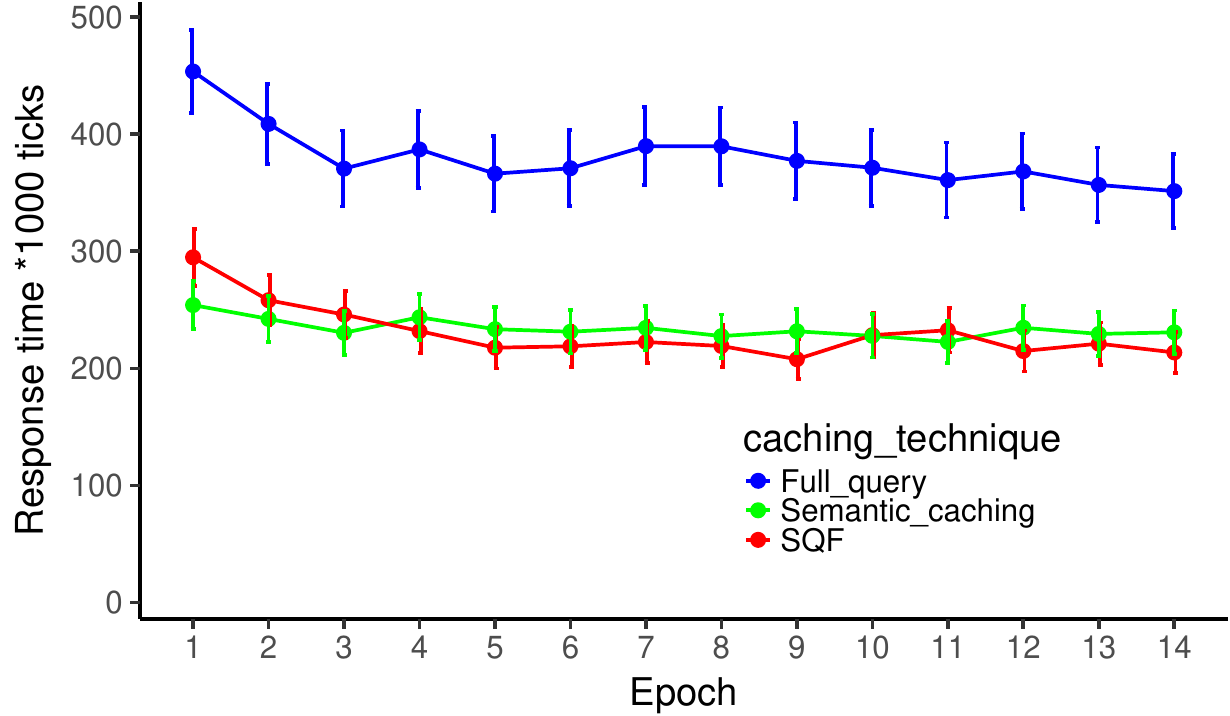}}
\end{minipage}\par\medskip
\centering
\subfloat[Uniform query input]{\label{fig:ResponseTime_Granularity_Uniform}\includegraphics[scale=.6]{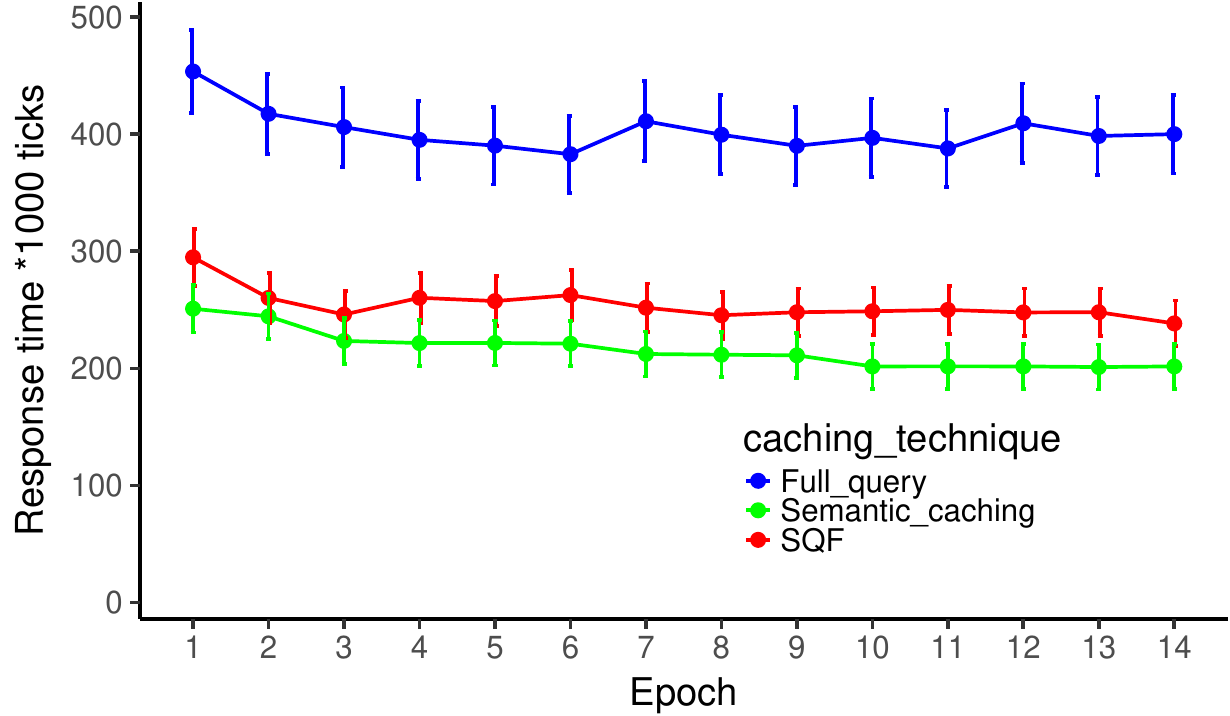}}

\caption{Response time for static workloads }
\label{fig:evaluation_comparison_Granularity}
\end{figure}

\subsubsection*{Discussion:}

The response time is observed for continuous epochs is plotted in  Figures \ref{fig:ResponseTime_Granularity_Poisson},  \ref{fig:ResponseTime_Granularity_Exponential}, and  \ref{fig:ResponseTime_Granularity_Uniform}. 
The full-query resultset caching model stores results as a single unit at one location. Hence it resulted in a high number of cache faults 
and high response time. In general, this policy  observes  high response time  for all types of query distributions. SQF and SC models performed almost similarly for Exponential and Poisson  distributions. Under similar conditions  SQF follows  semantic rules for the fragmentation of queries. The semantic model showed up to 10\% lower response times than SQF for Uniform distribution. One reason for SQF to have high response time could be that SQF  further fragments queries according to user requirements and association with other queries in  caches.  Since  sub-queries are repeated uniformly, the overall response time is slightly higher for SQF.

\subsection{Adaptivity to Changing Workloads}
\begin{figure}[ht]
\centering
\includegraphics[scale=0.8]{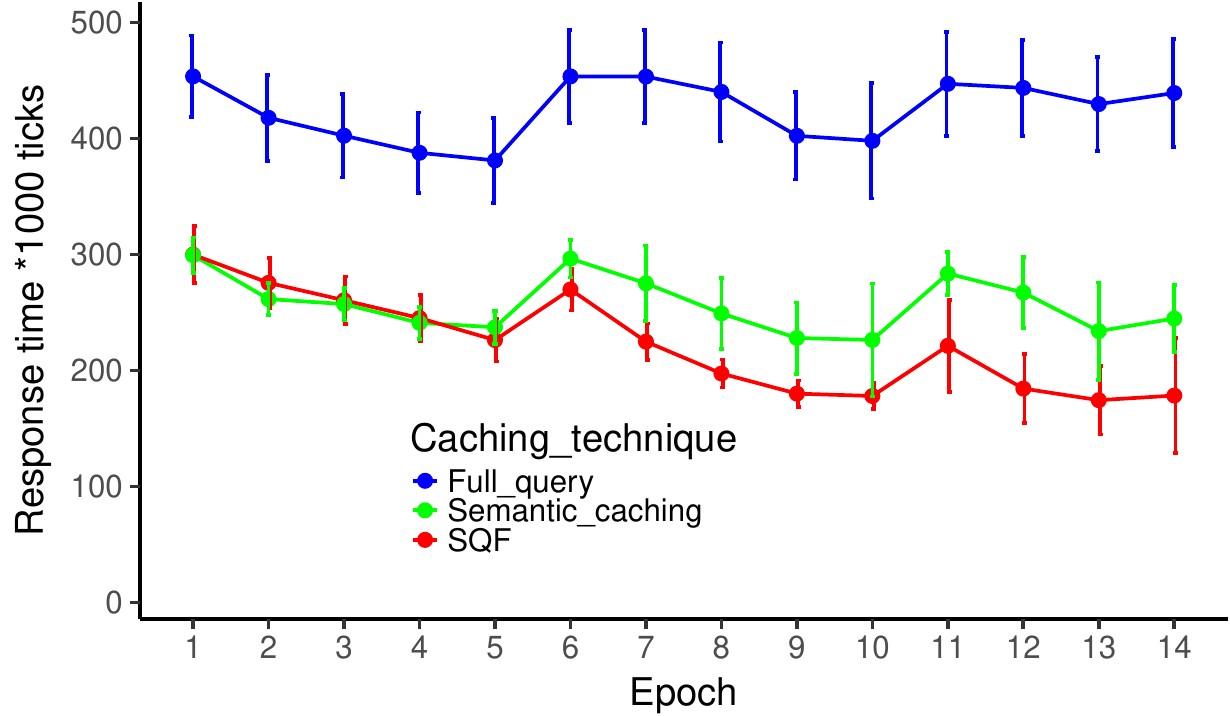}
\caption{Adaptivity to changing workloads}
\label{fig:adaptive_Uniform_10000_3}
\end{figure}

The plot in the Figure \ref{fig:adaptive_Uniform_10000_3} shows the adaptivity of caching techniques to changing query inputs over a continuous period. The input query distribution patterns are changed after  every five epochs. The experimental settings are similar to the above experiment. Query workloads are mixed  distribution of sub-query repetition.
\subsubsection*{Discussion:}
 A sudden spike in response time is observed after fifth and tenth epochs due to the change  introduced in the query input. Though, SQF and semantic models responded similarly for the first five epochs, as SQF caches highly associated fragments as a grain, it adapted much better to the variation in the workload than semantic caching or full query model. The advantage of SQF is even more evident with subsequent epochs. This shows that SQF is more powerful than other state-of-art techniques in the distributed environment where the environment is highly  dynamic and workloads are continuously changing as it happens in the real-world applications. 
	
\subsection{Cache Utilisation - Data Found Vs Caching Policies}
The percentage of data found in the cache influences  the response time. Higher volumes of data found in the cache ($\mathnormal{V}^{found}_{Q}$) reflect the ability of cache selection to maximise the cache utilisation.  
In this experiment, we allowed cache models to learn from query patterns. The learning is done by finding out the finest grain of query fragments that are repeated more than  a threshold. The learning gets the support from the associated sub-queries with each of the frequented sub-query. The percentage of data found in cache is compared across three caching policies.

\begin{figure}[ht]
\centering
\includegraphics[scale=0.8]{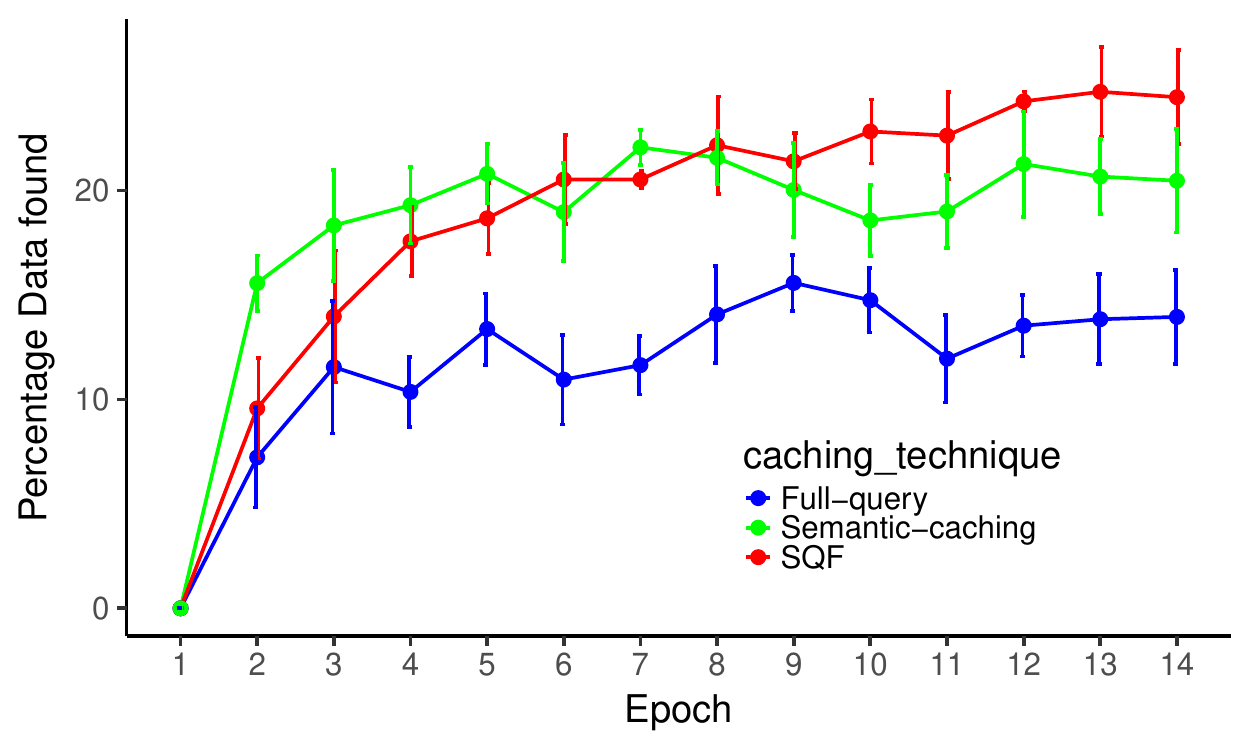}
\caption{Comparison percentage data found in cache across caching techniques}
\label{fig:PCDataFound_granularity_frequency_uniform}
\end{figure} 
\subsubsection*{Discussion:}
It is observed from the Figure \ref{fig:PCDataFound_granularity_frequency_uniform} that the semantic model performed better than other approaches during the initial epochs. Although initial performance of SQF in terms of percentage of data found in cache was lower than this of SC,   SQF observed the finer grains of query fragment repetition and cached the query fragments effectively. The observation capability provided SQF to ``learn'' user requirements up to the finest grain  and performed better over the subsequent epochs. Nearly 12.5\% more data is found than in the case of semantic caching model.  The results in Figure \ref{fig:PCDataFound_granularity_frequency_uniform} are presented for Uniform distribution. 

\subsection{Cache Stability - Inter-Cache Data Transfers}
\label{subsubsec:DT}
The  stability of a cache refers to the ability to predict future needs and cache appropriate data.  The  cache stability  measures the utilisation of cache. Higher utilisation of cache content refers to better performance. In the distributed cache environment, stability  is a representation of (i) fewer number of relocations of the cached data, (ii) high content re-use and (iii) higher prediction accuracy. Since query and data access patterns often exhibit locality of reference (temporal and spatial locality), data should be cached  near the user location where the data has been requested recently (for temporal locality) or stored close to the  location of the user (for spatial locality). Let, \\ \\
 Number of sub-query objects to relocate = $\delta_n$\\
 Average volume of a sub-query ($\mathnormal{q}_i$) in GB = $v_{qi}$\\
 Average data transfer cost  for inter-cache transfers per GB= $\mathnormal{d}_{net}$

\begin{equation}
\text{Average cost Inter-cache data transfers} = \frac{1}{\delta_n} \sum_{i=1}^{\delta_n} ({\delta_n} * v_{qi} * \mathnormal{d}_{net})
\end{equation}
The parameter ($\delta_n$) varies with the efficiency of the data placement algorithm  to place a data segment.   The parameter ($v_{qi}$) varies with  changing workloads. Queries with higher complexity lead to higher number of data accesses and hence transfers.   The following experiment  compares the response time   for inter-cache data transfers to reach to the cache set nearer to the user. The comparison is made using three cache techniques. (i) caching with sub-query fragmentation (SQF), (ii) semantically distributed data  and (iii) caching full-query resultsets.
 
 \begin{figure}[ht]
 \centering
 \includegraphics[scale=0.8]{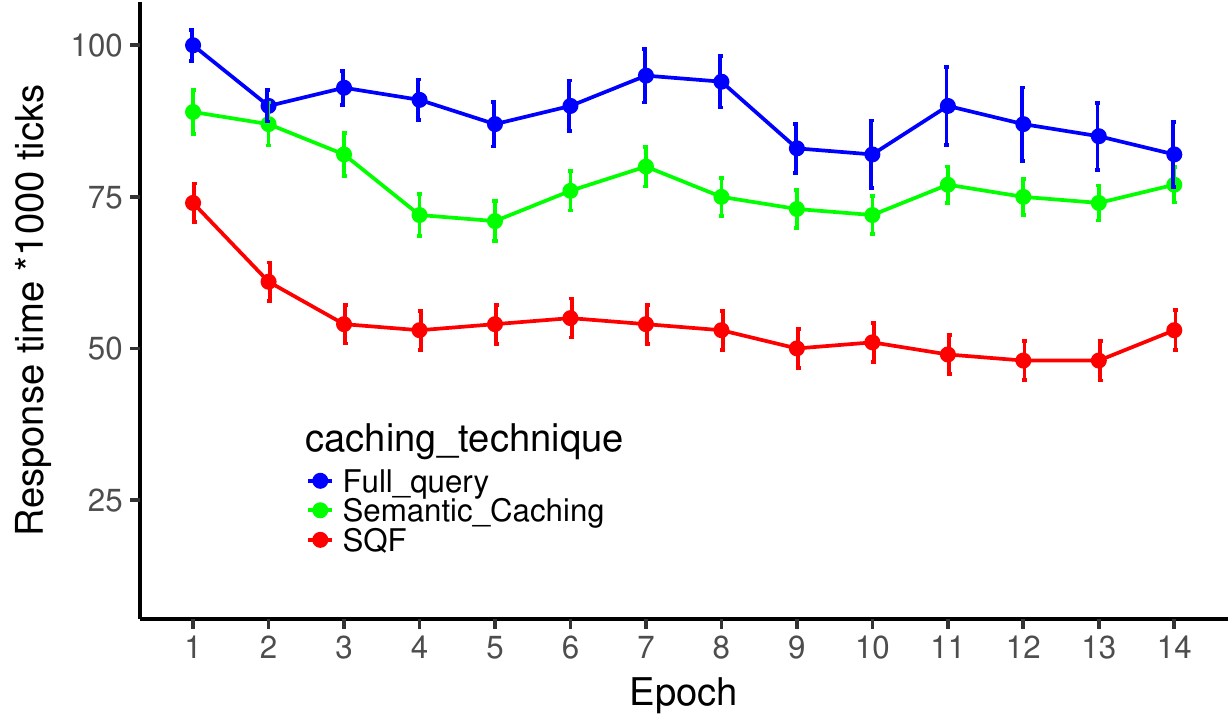}
 \caption{Response time due to inter-cache data transfers }
 \label{fig:PC_intercache_transfers_granularity_frequency_uniform}
 \end{figure}

\subsubsection*{Discussion:}
The results in Figure \ref{fig:PC_intercache_transfers_granularity_frequency_uniform}  presented represent Uniform distribution of sub-queries. It is observed that the  sub-query fragmentation (SQF) showed a clear advantage over other caching techniques used in  distributed environment. SQF has fewer inter-cache transfers as this approach stores all related data segments together. 

\subsection{Overhead of Sub-query Fragmentation}
Since sub-query fragmentation caches  more frequently repeated fragments of a query, it is often possible to cache the same data segment as a part of several other sub-queries before it is identified as a frequent fragment.  For example, a query $\mathcal{Q}$ has three sub plans : $\mathnormal{a}, \mathnormal{b}$ and $\mathnormal{c}$. Assume, after some time it is found that $\mathnormal{a}, \mathnormal{b}$ are as frequently queried together as $\mathnormal{a}, \mathnormal{c}$.  The SQF decides to cache $\mathnormal{ac}$ and $\mathnormal{ab}$ as two sub-queries. A repetition of data for the fragment $\mathnormal{a}$. Where as, semantic segment caches $\mathnormal{a}$ only once. 
So, SQF clearly tends to store more data initially. However, in the distributed environment, if the data is frequently needed at two locations at all times, then it makes sense to create copies of data and distribute them over different locations. This will have a  significant reduction in the query response time. Since $\mathnormal{a}$ is joined with $\mathnormal{b}$, the sub-query will be treated as an independent unit from $\mathnormal{ac}$ in SQF. Here we could not present all our findings and methods to identify the association between frequently queried data segments due to lack of space.

\section{Conclusion \& Future Work}
The work presented in this paper is a part of the research project - CommCache, a community shared cache framework for the optimization of data transfers. In this paper, we have discussed the problem of identifying suitable data fragments to cache in the distributed environment. The proposed sub-query fragmentation technique fragments user queries into sub-queries based on the repetition of partial query segments (sub-queries).  They  are stored together to provide quick retrieval of  data. Though SQF is an extension to the semantic caching, sub-queries are modeled  as  portable  objects suitable  to the distributed environment. 

Overall, the  SQF approach performed better than other caching approaches in the distributed environment. 
For the average response time, SQF is very effective in dynamic environments where workloads are changing as it is able to adapt to those changes quicker and better than existing approaches. Using traditional workloads, SQF outperformed other methods in most of the instances. It is only second best, after SC for static workloads.  Since SQF can be used find patterns of data accesses, it is possible to cache more accurate data than other techniques. Over a period of time, the percentage of data found is higher than  other methods. 

At present, SQF is restricted to the following limitations. The approach is mainly focused on distributed caches to understand the patterns of user queries using distributed learning. Hence  a centralised global query processor is assumed to manage the distribution of query segments. In future, we would like to extend the use of SQF for de-centralised environments. Another limitation is, the SQF approach depends on query execution  plans. This means, a different execution plan might lead to  reprocessing of the entire query. However, the query planner can consult existing cached contents before creating a plan. 

In future, we would like to extend the effectiveness of SQF in two directions: One, develop request-reply systems  from approximated cached results. It is to rank sub-queries in the order of priority and formulate reply with existing cached contents. Second, several query fragmentation techniques and  organisation of query execution are available for the execution of queries on shared memory and multi-core systems \cite{Lee2007, Lee2009}. It will be interesting to implement these techniques for distributed caches. We would like to pursue more study in this direction in future.

\bibliographystyle{unsrt}
\bibliography{FinalThesisDB.bib}

\end{document}